\documentclass[final,3p,times,number,sort&compress]{elsarticle}
\usepackage{amssymb}
\usepackage{amsthm}
\usepackage{amsmath}
\usepackage{url}
\usepackage{empheq}

\journal{}

\begin{document}

\begin{frontmatter}

%% Title, authors and addresses

%% use the tnoteref command within \title for footnotes;
%% use the tnotetext command for theassociated footnote;
%% use the fnref command within \author or \address for footnotes;
%% use the fntext command for theassociated footnote;
%% use the corref command within \author for corresponding author footnotes;
%% use the cortext command for theassociated footnote;
%% use the ead command for the email address,
%% and the form \ead[url] for the home page:
%% \title{Title\tnoteref{label1}}
%% \tnotetext[label1]{}
%% \author{Name\corref{cor1}\fnref{label2}}
%% \ead{email address}
%% \ead[url]{home page}
%% \fntext[label2]{}
%% \cortext[cor1]{}
%% \address{Address\fnref{label3}}
%% \fntext[label3]{}

\title{
System-size dependence of a jam-absorption driving strategy to remove traffic jam caused by a sag under the presence of traffic instability
}

%% use optional labels to link authors explicitly to addresses:
%% \author[label1,label2]{}
%% \address[label1]{}
%% \address[label2]{}

\author[label1]{Ryosuke Nishi\corref{cor1}}\ead{nishi@tottori-u.ac.jp}
\cortext[cor1]{Corresponding author. Tel.: +81 857 31 5192; fax: +81 857 31 5210.}
\address[label1]{Department of Mechanical and Physical Engineering, Faculty of Engineering, Tottori University, 4-101 Koyama-cho Minami, Tottori 680-8552, Japan.}
\author[label2]{Takashi Watanabe}
\address[label2]{Department of Engineering, Graduate School of Sustainability Science, Tottori University, 4-101 Koyama-cho Minami, Tottori 680-8552, Japan.}

%arXiv: abstracts longer than 1920 characters will not be accepted
% 1531 characters
\begin{abstract}
Sag is a road section where a downhill changes into an uphill, and is a highway bottleneck. We consider a system in which all vehicles are connected, and run on a single-lane road with a sag. We propose a simple strategy for removing each traffic jam caused by the sag. Our strategy assigns a vehicle upstream of the jam front to perform the jam-absorption driving (JAD): running toward the predicted goal, and finally removing the jam. We use a microscopic car-following model possessing the traffic instability, an acceleration model against the road gradient of a sag, and an instantaneous fuel consumption model. Our main goal is to elucidate the influence of the system size (the number of vehicles in the system) on our strategy. By increasing the system size from 500 to 10000 vehicles, we have found the following results for the average total travel time per vehicle, and the average total fuel consumption per vehicle. Our strategy can reduce the former with a slightly increasing rate of reduction. Our strategy can reduce the latter with a rate of reduction which decreases and becomes roughly constant. Optimal spatiotemporal scales of JAD for the former and the latter become roughly constant, respectively. Minimizing the former and the latter simultaneously is not possible. Not only vehicular traffic flow, but also the collective dynamics of other self-driven particles (such as ships, swarm robots, and pedestrians) lined up in a single column, and passing through a bottleneck can be modeled using our strategy.
\end{abstract}


\end{frontmatter}

%% \linenumbers

%% main text
%%%%%%%%%%%%%%%%%%%%%%%%%%%%%%%%%%%%%%%%%%%%%%%%%%%%%%%%%%%%%%%%%%%%%
\section{\label{sec:introduction}Introduction}
\begin{figure}[t]
\centering
\includegraphics[width=\hsize]{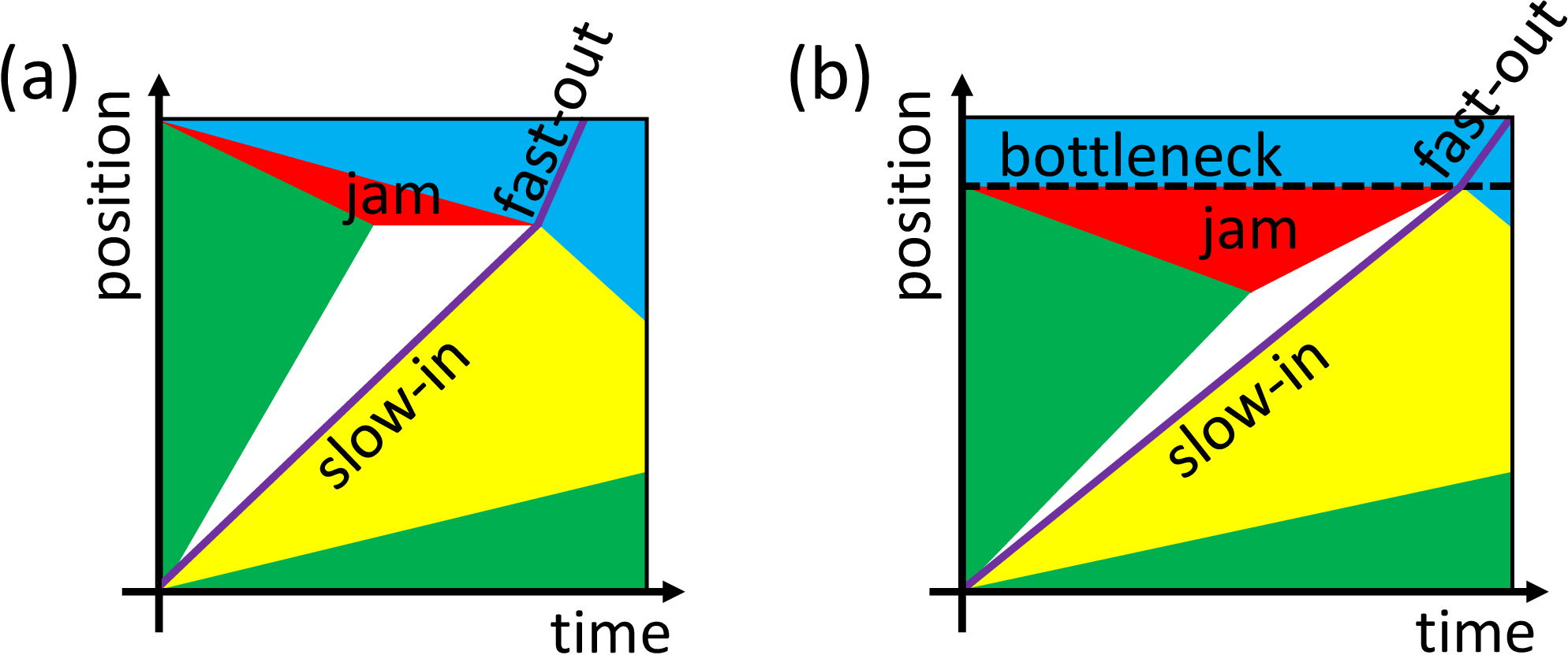}
\caption{
Schematic views of (a) the traffic jam whose downstream front propagates upstream (the wide moving jam), and (b) the traffic jam whose downstream front is fixed at a bottleneck. Each traffic jam is removed by the jam-absorption driving (JAD) composed of the slow-in and the fast-out phases.
Red regions denote the states inside the traffic jams.
Green and blue regions denote the states upstream and downsteam of the traffic jams, respectively.
Yellow regions denote the states influenced by the slow-in phase.
Purple solid lines denote the trajectories of the absorbing vehicles. Black dashed line denotes the bottleneck.
}
\label{fig:JAD_schematic_view}
\end{figure}

%%%%%%%%%%%%%%%%%%%%%%%%%%%%%%%%%%%%%%%%%%%%%%%%%%%%%%%%%%%%%%%%%%%%%
Highway traffic flow has been studied extensively. Collective dynamics of vehicles in highways has been revealed as a kind of the dynamics of self-driven particles~\citep{Chowdhury2000, Helbing2001, Kerner2004Physics, Schadschneider2010Stochastic, Treiber2013}.
Additionally, strategies based on ramp metering (RM)~\citep{Papageorgiou2002} and variable speed limit (VSL)~\citep{Lu2014, Khondaker2015}, and using essentially road-side infrastructures have been developed to improve highway traffic flow.
For instance, a dynamic VSL strategy known as the ``SPEed ControllIng ALgorIthm using Shockwave Theory (SPECIALIST)" removes a wide moving jam~\citep{Hegyi2008,Hegyi2010}.
A dynamic strategy known as the ``mainstream traffic flow control" blocks the occurrence of the capacity drop by limiting the mainstream flow upstream of a bottleneck with RM and/or VSL~\citep{Carlson2010a,Carlson2010b}.
Dynamic VSL strategies also improve the flow rate at a bottleneck~\citep{Chen2014}, and mitigate traffic jam with distributed~\citep{Popov2008} or model predictive~\citep{Hegyi2005a, Han2017a} control algorithms.
Moreover, strategies have been developed which directly manipulate the spatiotemporal maneuvers of special vehicles to improve the traffic flow on highways and related roads, and some of them utilize connected and/or automated vehicles~\citep{Vahidi2018,Wang2020c,Yu2021}.
Strategies in this direction include basic concepts~\citep{Beaty1998,Washino2003}, Pacer Cars~\citep{Behl2010}, and the jam-absorption driving (JAD)~\citep{Nishi2013}.
JAD is described as a chain of two actions \textit{slow-in} and \textit{fast-out} performed by a single vehicle (called the \textit{absorbing vehicle}) to remove traffic jam as shown in Figs.~\ref{fig:JAD_schematic_view}(a) and (b). During the slow-in phase, the absorbing vehicle, which is sufficiently upstream of the jam, stops following the preceding vehicle, and decelerates to a lower velocity resulting in a large gap between it and its preceding vehicle. Owing to this enlarged gap, the absorbing vehicle avoids getting caught in the jam, and the jam shrinks and finally disappears. Subsequently, the absorbing vehicle promptly returns to following the preceding vehicle as the fast-out phase.
The performance of JAD has been investigated numerically and/or theoretically using kinematic~\citep{Nishi2013} and car-following~\citep{Taniguchi2015, He2017, Zheng2020, Nishi2020, Wang2021} models, and experimentally using real vehicles on a circuit~\citep{Taniguchi2015TGF}.
Strategies in this direction has been studied diligently~\citep{Jerath2015,Wu2021a,Stern2018,Zheng2020IEEE,Li2020,Dayi2021,Han2021,Han2021a}, and has been also developed as the Lagrangian control of moving bottlenecks~\citep{Ramadan2017,Piacentini2018a,Cicic2018,Yang2019b}, and the speed harmonization~\citep{Ma2016,Yang2017e,Learn2018a,Guo2021}.
Additionally, strategies have been developed which change the car-following performance of automated vehicles (AVs)~\citep{Kesting2008}, and connected vehicles (CVs)~\citep{Knorr2012} according to the traffic situation.
A strategy combining VSL, RM, and CVs with a penetration rate of $100\,\%$ is helpful in clearing traffic jam~\citep{VandeWeg2014}. Trajectory planning has also been developed for connected and automated vehicles~\citep{Li2018TRB,Li2019}.
It should be noted that strategies manipulating spatiotemporal maneuvers of vehicles have also been studied in city traffic (such as the traffic at signalized arterials and intersections)~\citep{Vahidi2018}. This study focuses on highway traffic.
%%%%%%%%%%%%%%%%%%%%%%%%%%%%%
%%%%%%%%%%%%%%%%%%%%%%%%%%%%%

Highway traffic jam is categorized into two types~\citep{Kerner2004Physics}: (i) the traffic jam whose downstream front propagates upstream as shown in Fig.~\ref{fig:JAD_schematic_view}(a) (this type of traffic jam is called the wide moving jam), and (ii) the traffic jam whose downstream front is fixed at a bottleneck as shown in Fig.~\ref{fig:JAD_schematic_view}(b). For both types of traffic jam, upstream tails propagate upstream.
In this study, we focus on the second type as the target traffic jam, and consider a bottleneck on a highway road.
Researchers have treated various bottleneck types or effects:
fixed bottleneck (such as traffic accidents, work zones, and obstacles)~\citep{Ramadan2017,Piacentini2018a,Piacentini2019a,Ko2020,Cicic2021},
lane reduction~\citep{Li2018TRB,Vinitsky2018a,Ha2020a},
on-ramp~\citep{Kesting2008,Knorr2012,Cicic2021,Guo2021},
rubbernecking in a specific region~\citep{He2017},
imposition of capacity drop~\citep{Han2017,Ghiasi2019,Nishi2020},
and sag~\citep{Morino2016,Watanabe2018a,Goni-Ros2016,Nezafat2018TRR}.
Among various bottlenecks on highways, this study focuses on sag. Sag is a section of the road where a gentle downhill changes into an uphill, and is known as a major bottleneck. Traffic jam occurs in a sag because vehicles tend to decelerate in the uphill, and the deceleration is amplified along vehicle strings.

Various strategies manipulate special vehicles to improve highway traffic at bottlenecks.
Examples of them include
trajectory planning~\citep{Li2018TRB},
changing the settings of adaptive cruise control according to the traffic situation~\citep{Kesting2008},
running slowly upstream of traffic jam~\citep{Knorr2012,Ramadan2017,Piacentini2018a,Cicic2021,Morino2016,Watanabe2018a,Nezafat2018TRR},
maneuvers combining deceleration and acceleration~\citep{Goni-Ros2016},
maneuvers produced by reinforcement learning~\citep{Vinitsky2018a,Ha2020a},
reinforcement learning of merge control and speed harmonization~\citep{Ko2020},
combination of cooperative adaptive cruise control platooning, cooperative merge, and speed harmonization~\citep{Guo2021},
and the maneuvers aiming to remove the traffic jam such as Pacer Cars~\citep{Behl2010}, speed harminization~\citep{Ghiasi2019}, dynamic VSL with CVs~\citep{Han2017}, and JAD~\citep{He2017,Nishi2020}.
Among these strategies, this study focuses on JAD.

When an absorbing vehicle performs JAD to clear a traffic jam, its actions cause perturbations to the upstream traffic flow. Such a perturbation may grow into new traffic jam (also known as the secondary traffic jam)~\citep{Taniguchi2015,He2017,Nishi2020,Zheng2020,Han2021a}. The secondary traffic jam has been also reported and investigated in the SPECIALIST and related strategies~\citep{Hegyi2010,Wang2012eIJMPC,Wang2014fIJMPC,Wang2016}.
As the system size (the number of vehicles in a system) increases, secondary traffic jam may propagate further upstream and deteriorate the performance of JAD in terms of the total travel time and the total fuel consumption.
Therefore, we need to investigate the influence of the system size on JAD strategies for highway bottlenecks.
It should be noted that we consider the influence of the system size solely, and do not consider the influence of the initial density of vehicle platoons (${\rm veh}/{\rm km}$), traffic demand (${\rm veh}/{\rm h}$), or road length.
Han et al.~\citep{Han2017} evaluated the influence of the number of vehicles (or the duration of high traffic demand) on the performance of their CV-based VSL strategies. They used a kinematic traffic flow model, set directly the probability that each vehicle causes traffic instability at a bottleneck, and calculated the probability of traffic breakdown when a given number of vehicles pass through the bottleneck.
However, since the kinematic model they used did not possess the traffic instability, they did not consider the secondary traffic jam caused by the actions of the special vehicles, and arising upstream of the bottleneck. To treat such secondary traffic jam, we need to use the traffic flow models inherently possessing the traffic instability.
Thus, the aim of this study is to elucidate the system-size dependence of a JAD strategy to remove traffic jam caused by a sag under the presence of traffic instability.

In this study, we consider a system in which a given number of vehicles ranging from 500 to 10000 vehicles run on a non-periodic and single-lane highway road with a sag, and all vehicles are CVs.
We design a JAD strategy in this system. We detect the occurrence of the target traffic jam fixed at the sag by collecting the current positions and velocities of the CVs, and predict the future time-space points of the downstream front of the traffic jam (DFTJ). Next, we assign a CV, which is the nearest CV upstream of the DFTJ by a preset distance or more, to the absorbing vehicle. In the slow-in phase, the absorbing vehicle stops following its preceding vehicle, and moves toward the goal, which is near the future time-space point where the DFTJ disappears. We repeatedly update the current DFTJ, the goal, and the maneuver of the absorbing vehicle. Once the absorbing vehicle reaches the goal position, the jam gets cleared, and then the absorbing vehicle finishes its slow-in phase and restarts following its preceding vehicle as the fast-out phase. Since the sag generates the next traffic jam, we try to assign the next absorbing vehicle to remove the next jam just after an absorbing vehicle completes its slow-in phase.

We evaluate the performance of our JAD strategy by numerical simulations.
We represent microscopic car-following behaviors of vehicles using the intelligent driver model plus (IDM+)~\citep{Schakel2013}, which is a microscopic car-following model modified from the intelligent driver model (IDM)~\citep{Treiber2000}. The IDM+ possesses traffic instability, and has been used widely~\citep{Goni-Ros2016,Morino2016,Watanabe2018a,Nezafat2018TRR,Goni-Ros2019}.
Additionally, we reproduce the longitudinal movement of vehicles in a sag using a model of the acceleration of vehicles against the road gradient of a sag proposed by Go\~{n}i-Ros et al.~\citep{GoniRos2016IJITSR}. This model has been used for the numerical simulations of the traffic flow in a sag~\citep{Goni-Ros2016,Nezafat2018TRR,Goni-Ros2019}.
As the performance indicators of our JAD strategy, we evaluate the total travel time, the total fuel consumption, and average values of them per vehicle.
To estimate fuel consumption, we use the emissions from traffic (EMIT)~\citep{Cappiello2002} as a typical model of instantaneous fuel consumption and emissions, which has been demonstrated to be suitable for estimating fuel consumption~\citep{Guo2015}.

The remainder of this paper is organized as follows.
We describe the modeling of the system in Sec.~\ref{sec:system}, design the traffic scenarios and our JAD strategy in Sec.~\ref{sec:scenarios}, define the performance indicators in Sec.~\ref{sec:performance_indicators}, and present the results in Sec.~\ref{sec:results}. Section~\ref{sec:discussion} describes the conclusive discussion.
%%%%%%%%%%%%%%%%%%%%%%%%%%%%%%%%%%%%%%%%%%%%%%%%%%%%%%%%%%%%%%%%%%%%%
\section{\label{sec:system}System}
%%%%%%%%%%%%%%%%%%%%%%%%%%%%%%%%%%%%%%%%%%%%%%%%%%%%%%%%%%%%%%%%%%%%%
%%%%%%%%%%%%%%%%%%%%%%%%%%%%%%%%%%%%
\begin{figure}[t]
\centering
\includegraphics[width=\hsize]{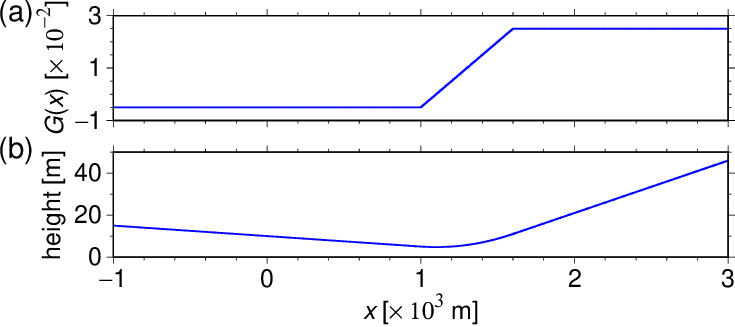}
\caption{
(a) Road gradient distribution. (b) Road height distribution. We set the road gradient distribution according to Refs.~\citep{GoniRos2016IJITSR,Goni-Ros2016,Goni-Ros2019}, and the road height at $x = 1000$ to $5\,{\rm m}$.
}
\label{fig:sag}
\end{figure}
%%%%%%%%%%%%%%%%%%%%%%%%%%%%%%%%%%%%
We consider an infinitely long, non-periodic, and single-lane highway road with a sag. This road does not have any curves, on-ramps, or off-ramps.
Parameter $x$ (m) is the longitudinal position on this road, and $G(x)$ is the road gradient at position $x$.
The road gradient $G(x)$ is positive for the uphill terrain. To avoid confusion, we emphasize that $G(x)$ does not denote a percentage value throughout this study. For instance, $G(x) = 0.01$ denotes the road gradient at position $x$ corresponding to an increase of $1\,{\rm m}$ by running horizontally for $100\,{\rm m}$, and not $10000\,{\rm m}$.
According to Refs.~\citep{GoniRos2016IJITSR,Goni-Ros2016,Goni-Ros2019}, we set a road gradient distribution for this road which resembles the road gradient distribution for the Yamato Sag, Tomei Expressway, Japan~\citep{Patire2011} as follows:
\begin{subequations}
\begin{empheq}[left={G(x) = \empheqlbrace \,}]{alignat=2}
& -0.005 & \quad \mbox{if $x \le 1000$,} \label{eq:G_x_a} \\
& -0.005 + 5 (x - 1000) \times 10^{-5} & \quad \mbox{if $1000 \le x \le 1600$,} \label{eq:G_x_b} \\
& 0.025 & \quad \mbox{if $x \ge 1600$,} \label{eq:G_x_c}
\end{empheq}
\label{eq:G_x}
\end{subequations}
where the first section ($x \le 1000$) is the downhill, the second section ($1000 \le x \le 1600$) is the sag vertical curve, and the third section ($x \ge 1600$) is the uphill.
Figures~\ref{fig:sag}(a) and (b) indicate the road gradient distribution and an example of the road height distribution for this road, respectively.

We initially place $N$ vehicles on the road in each run. They do not disappear from the road, and no other vehicles do not appear on the road in the middle of each run. Hence, the number of vehicles $N$ is fixed throughout each run. Vehicle 1 is the leading vehicle, vehicle $i$ is located just behind vehicle $i-1$ ($i = 2, \ldots, N$), and vehicle $N$ is the last vehicle. The vehicles are not allowed to overtake those ahead of them. We define $x_i(t)$ (m), $v_i(t)$ (${\rm m}/{\rm s}$) and $a_i(t)$ (${\rm m}/{\rm s^2}$) as the front-edge position, velocity, and acceleration of vehicle $i$ at time $t$ (s), respectively.
Each vehicle in the system is a CV, and its current position and velocity are detected with an on-board GPS device and shared with the other vehicles via vehicle-to-vehicle (V2V) and vehicle-to-infrastructure (V2I) communication.
For simplicity, we assume that each vehicle obtains the current positions and velocities of all vehicles without any data loss, noise, or delays.
We also assume that each vehicle is a human-driven vehicle, and not an AV whose velocity is controlled automatically.

In our numerical simulations, each run starts at the initial time $t = 0\,{\rm s}$, and ends when the last vehicle reaches $x=6000\,{\rm m}$. Since no boundary exists on the road, vehicles keep moving downstream untill the end of each run.
We update $x_i(t)$ and $v_i(t)$ at regular time intervals of $\Delta t = 0.1\,{\rm s}$ by using the ballistic method as an efficient numerical integration scheme for vehicular traffic flow~\citep{Treiber2015}. The below equations are used to update the values of $x_i(t)$ and $v_i(t)$:
\begin{align}
x_i(t + \Delta t) &= x_i(t) + v_i(t) \Delta t + \dfrac{a_i(t) \Delta t^2}{2}, \label{eq:x_update_ballistic} \\
v_i(t + \Delta t) &= v_i(t) + a_i(t) \Delta t. \label{eq:v_update_ballistic}
\end{align}
It should be noted that we consider only the horizontal components of vehicles, and not the vertical components according to Refs.~\citep{Goni-Ros2016,Goni-Ros2019}. We believe that this simplification does not deteriorate the performance of our numerical simulations because the absolute value of the road gradient is considerably small ($|G(x)| \le 0.025$).

Vehicle $i$'s acceleration $a_i(t)$ is given by~\citep{Goni-Ros2016}
\begin{align}
a_i(t) = \max \left\{ a_{{\rm des},\,i}(t) + \delta_i(t), a_{\rm min}, -\dfrac{v_i(t)}{\Delta t} \right\},
\label{eq:a_i_t}
\end{align}
where $a_{{\rm des},\,i}(t)$ (${\rm m}/{\rm s^2}$) is the desired acceleration of vehicle $i$ at time $t$,
$\delta_i(t)$ (${\rm m}/{\rm s^2}$) is the effect of the road gradient on the acceleration of vehicle $i$ at time $t$,
$a_{\rm min}$ (${\rm m}/{\rm s^2}$) is the minimum acceleration,
and $-v_i(t)/\Delta t$ is the acceleration of vehicle $i$ to stop at the next time $t + \Delta t$ and is used for preventing negative velocities.
We set $a_{\rm min} = -8\,{\rm m}/{\rm s^2}$~\citep{Goni-Ros2016} as listed in Table~\ref{table:param}.

According to the IDM+~\citep{Schakel2013}, the desired acceleration $a_{{\rm des},\,i}(t)$ is given by
\begin{align}
a_{{\rm des},\,i}(t) = a \min \left\{ 1 - \left( \dfrac{v_i(t)}{v_0} \right)^{\delta}, \, 1 - \left( \dfrac{s^{\ast} \left( v_i(t), \Delta v_i(t) \right) }{s_i(t)} \right)^2 \right\}.
\label{eq:a_des_i_t_IDM_plus}
\end{align}
Variable $s_{i}(t)$ (m) is the gap between the rear edge of vehicle $i-1$ and the front edge of vehicle $i$ at time $t$, and is given by
\begin{align}
s_i(t) = x_{i-1}(t) - x_i(t) - d,
\label{eq:s_i_t}
\end{align}
where $d$ (m) is the vehicle length. We set $d = 4.5\,{\rm m}$ on the basis of Ref.~\citep{Cappiello2002}, and list the value in Table~\ref{table:param} (see Sec.~\ref{sec:performance_indicators} for the reason for using this value).
Variable $\Delta v_i(t)$ (${\rm m}/{\rm s}$) is the relative velocity between vehicles $i$ and $i-1$ at time $t$, and is given by
\begin{align}
\Delta v_i(t) = v_i(t) - v_{i-1}(t).
\label{eq:Delta_v_i_t}
\end{align}
Variable $s^{\ast}(v_i(t), \Delta v_i(t))$ (m) is the desired gap of vehicle $i$ at time $t$, and is given by~\citep{Treiber2013}
\begin{align}
s^{\ast} \left( v_i(t), \Delta v_i(t) \right) = s_0 + \max \left\{ 0, T v_i(t) + \dfrac{v_i(t) \Delta v_i(t)}{2 \sqrt{ab}} \right\}.
\label{eq:s_asterisk}
\end{align}
Parameter $a$ (${\rm m}/{\rm s}^2$) is the maximum acceleration, $b$ (${\rm m}/{\rm s}^2$) is the comfortable deceleration, $s_0$ (m) is the gap in the halting state, $v_0$ (${\rm m}/{\rm s}$) is the desired velocity, $T$ (s) is the safe time gap, and $\delta$ is the exponent. We set these parameters according to Ref.~\citep{Goni-Ros2019}, and list them in Table~\ref{table:param}.
It should be noted that inter-driver~\citep{Ossen2011} and intra-driver~\citep{Wang2010,Wagner2012,Laval2014a,Taylor2015,Jiang2014,Huang2018c} heterogeneities in car-following behaviors have significant effect on vehicular traffic flow. Nevertheless, to compare different traffic scenarios easily, we ensure that the IDM+ parameters are constant and common to all vehicles~\citep{Goni-Ros2016}.
%%%%%%%%%%%%%%%%%%%%%%%%%%%%%%%
\begin{table}[t]
\caption{\label{table:param}
Parameter settings. We set the IDM+ parameters according to Ref.~\citep{Goni-Ros2019}, the parameters of the road gradient's effect according to Refs.~\citep{Goni-Ros2016,Goni-Ros2019}, and the EMIT parameters and the vehicle length according to Ref.~\citep{Cappiello2002}.
}
\begin{center}
\begin{tabular}{lr||lr||lr}
\hline
parameter (unit) & value & parameter (unit) & value & parameter (unit) & value \\
\hline
$a\,({\rm m}/{\rm s}^2)$          & 1.4    & $B\,({\rm kW}/({\rm m}/{\rm s})^2)$                       & $2.7384 \times 10^{-3}$ & $x_{\rm esc}\,({\rm m})$                 & 1600 \\
$b\,({\rm m}/{\rm s}^2)$          & 2.1    & $C\,({\rm kW}/({\rm m}/{\rm s})^3)$                       & $1.0843 \times 10^{-3}$ & $v_{\rm cau}\,({\rm m}/{\rm s})$         & 15   \\
$s_0\,({\rm m})$                  & 3      & $M_{\rm ton}\,(10^3 {\rm kg})$                            & 1.325                   & $v_{\rm esc}\,({\rm m}/{\rm s})$         & 28   \\
$v_0\,({\rm m}/{\rm s})$          & 30.56  & $g\,({\rm m}/{\rm s}^2)$                                  & 9.81                    & $h_{\rm R,\,pre}\,({\rm s})$             & 2    \\
$T\,({\rm s})$                    & 1.3    & $\alpha_{\rm FR}\,({\rm g}/{\rm s})$                      & 0.365                   & $h_{\rm R,\,min}\,({\rm s})$             & 1.3  \\
$\delta$                          & 4      & $\beta_{\rm FR}\,({\rm g}/{\rm s}/({\rm km}/{\rm h}))$    & 0.00114                 & $h_{\rm R,\,max}\,({\rm s})$             & 2.5  \\
$d\,({\rm m})$                    & 4.5    & $\delta_{\rm FR}\,({\rm g}/{\rm s}/({\rm km}/{\rm h})^3)$ & $9.65 \times 10^{-7}$   & $a_{\rm min,\,JAD}\,({\rm m}/{\rm s}^2)$ & $-1$ \\
$\lambda\,({\rm s}^{-1})$         & 0.0004 & $\zeta_{\rm FR}\,({\rm g}/{\rm s}/({\rm m}^2/{\rm s}^3))$ & 0.0943                  & $a_{\rm max,\,JAD}\,({\rm m}/{\rm s}^2)$ & 1    \\
$\theta\,({\rm m}/{\rm s}^2)$     & 22     & $\alpha'_{\rm FR}\,({\rm g}/{\rm s})$                     & 0.299                   & $a_{\rm min}\,({\rm m}/{\rm s}^2)$       & $-8$ \\
$A\,({\rm kW}/({\rm m}/{\rm s}))$ & 0.1326 & $x_{\rm end}\,({\rm m})$                                  & 5000                    &                                          &      \\
\hline
\end{tabular}
\end{center}
\end{table}
%%%%%%%%%%%%%%%%%%%%%%%%%%%%%%%

According to Ref.~\citep{GoniRos2016IJITSR}, acceleration $\delta_i(t)$ is determined as follows.
We assume that drivers compensate the road gradient gradually, if not instantly, and define $G_{{\rm com},\,i}(t)$ as the amount of compensated road gradient by the driver of vehicle $i$ at time $t$. We update $G_{{\rm com},\,i}(t)$ by the following equation:
\begin{subequations}
\begin{empheq}[left={G_{{\rm com},\,i}(t + \Delta t) = \empheqlbrace \,}]{alignat=2}
& G\left(x_i(t + \Delta t)\right)         & \quad \mbox{if $G\left(x_i(t + \Delta t)\right) \le G_{{\rm com},\,i}(t) + \lambda \Delta t$,} \label{eq:G_com_i_update_a} \\
& G_{{\rm com},\,i}(t) + \lambda \Delta t & \quad \mbox{if $G\left(x_i(t + \Delta t)\right) >   G_{{\rm com},\,i}(t) + \lambda \Delta t$.} \label{eq:G_com_i_update_b}
\end{empheq}
\label{eq:G_com_i_update}
\end{subequations}
Equation~(\ref{eq:G_com_i_update}) denotes that the driver of vehicle $i$ compensates for an increase in the road gradient with a maximum compensation rate $\lambda$ (${\rm s^{-1}}$). Once the driver has completed his/her compensation, the amount of compensated road gradient becomes equal to the actual road gradient. As $\lambda$ becomes larger and smaller, $G_{{\rm com},\,i}(t)$ moves toward the actual road gradient more quickly and slowly, respectively.
Equation~(\ref{eq:G_com_i_update}) also denotes that the driver instantly compensates for any decrease in the road gradient.
Using $G_{{\rm com},\,i}(t)$, we express
\begin{align}
\delta_i(t) = -\theta \left\{ G(x_i(t)) - G_{{\rm com},\,i}(t) \right\},
\label{eq:delta_i_t}
\end{align}
where $\theta$ (${\rm m}/{\rm s^2}$) is the sensitivity parameter, and is generally positive.
If the driver of vehicle $i$ does not fully compensate for the increased road gradient ($G_{{\rm com},\,i}(t) < G(x_i(t))$), $\delta_i(t)$ is negative. If the driver fully compensates ($G_{{\rm com},\,i}(t) = G(x_i(t))$), $\delta_i(t)$ is zero.
We set $\lambda = 0.0004\,{\rm s^{-1}}$ and $\theta = 22\,{\rm m}/{\rm s^2}$~\citep{Goni-Ros2016,Goni-Ros2019} as listed in Table~\ref{table:param}.

\begin{figure}[t]
\centering
\includegraphics[width=\hsize]{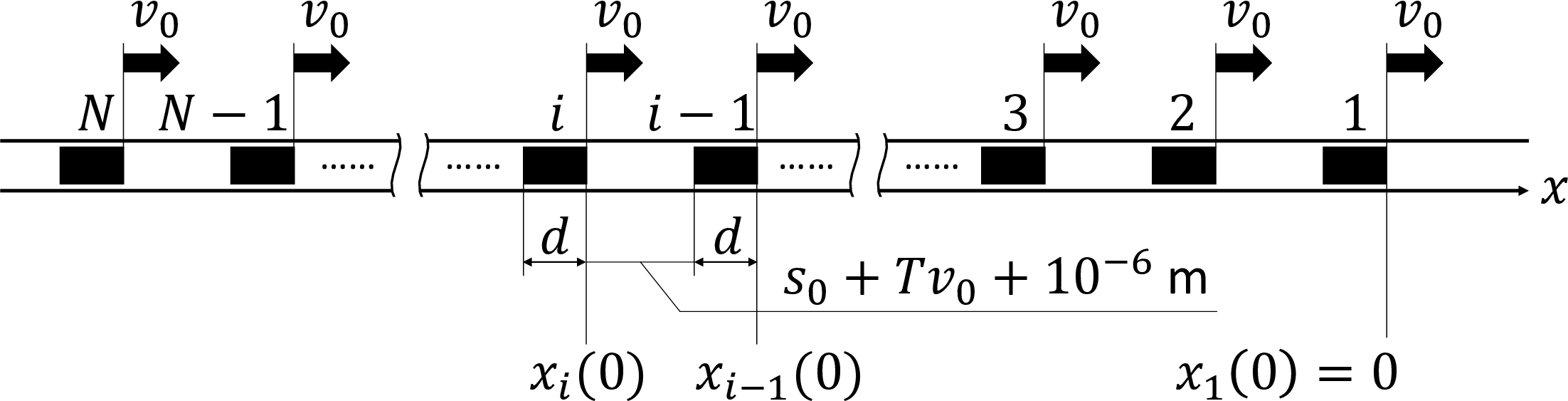}
\caption{
Initial conditions.
Vehicle 1's front edge is placed at $x = 0$.
%%% first submission start %%%
% Vehicles $2, 3, \ldots, N$ have the same gap of $s_0 + T v_0$.
%%% first submission end %%%
%%% revised start %%%
Vehicles $2, 3, \ldots, N$ have the same gap of $s_0 + T v_0 + 10^{-6}\,{\rm m}$.
%%% revised end %%%
Hence, all vehicles are placed downhill.
All vehicles have the same velocity of $v_0$ and the same acceleration of zero.
}
\label{fig:ini_conds}
\end{figure}

The initial conditions are presented in Fig.~\ref{fig:ini_conds}. At the initial time $t = 0\,{\rm s}$, we set the position of the front edge of the leading vehicle (vehicle 1) at $x = 0\,{\rm m}$,
%%% first submission start %%%
% and place the other vehicles with the same gap of $s_0 + T v_0$, which is the minimum equilibrium gap for running at the maximum velocity $v_0$, and set the velocities of all vehicles to $v_0$.
%%% first submission end %%%
%%% revised start %%%
place the other vehicles with the same gap of $s_0 + T v_0 + 10^{-6}\,{\rm m}$, and set the velocities of all vehicles to $v_0$.
Gap $s_0 + T v_0$ is the minimum equilibrium gap for running at the maximum velocity $v_0$. We do not set the initial gap to $s_0 + T v_0$ but set it to the sum of $s_0 + T v_0$ and a small gap $10^{-6}\,{\rm m}$ for making the initial flow linearly string stable (see Sec.~\ref{subsec:stability}).
%%% revised end %%%
That is, the initial position and velocity of vehicle $i$ ($i = 1, \ldots, N$) are given by
%%% first submission start %%%
% \begin{align}
% x_i(0) = -i ( d + s_0 + T v_0 )
% \label{eq:x_i_0}
% \end{align}
%%% first submission end %%%
%%% revised start %%%
\begin{align}
x_i(0) = -i ( d + s_0 + T v_0 + 10^{-6} )
\label{eq:x_i_0}
\end{align}
%%% revised end %%%
and
\begin{align}
v_i(0) = v_0,
\label{eq:v_i_0}
\end{align}
respectively.
Besides, the initial desired acceleration is set to zero for all vehicles:
\begin{align}
a_{{\rm des},\,i}(0) = 0.
\label{eq:a_des_i_0}
\end{align}
Since all vehicles are initially placed downhill, the road gradient does not influence the acceleration of vehicles at the initial time. Therefore, the initial values of $G_{{\rm com},\,i}(t)$ and $\delta_i(t)$ are given by
\begin{align}
G_{{\rm com},\,i}(0) = G(x_i(0)) = -0.005
\label{eq:G_com_i_0}
\end{align}
and
\begin{align}
\delta_i(0) = 0,
\label{eq:delta_i_0}
\end{align}
respectively for all vehicles ($i = 1, \ldots, N$).
%%%%%%%%%%%%%%%%%%%%%%%%%%%%%%%%%%%%%%%%%%%%%
%%%%%%%%%%%%%%%%%%%%%%%%%%%%%%%%%%%%%%%%%%%%%
%%%%%%%%%%%%%%%%%%%%%%%%%%%%%%%%%%%%%%%%%%%%%%%%%%%%%%%%%%%%%%%%%%%%%%%%%%%%%%%%%%%%%%%%%
\section{\label{sec:scenarios}Traffic Scenarios}
%%%%%%%%%%%%%%%%%%%%%%%%%%%%%%%%%%%%%%%%%%%%%%%%%%%%%%%%%%%%%%%%%%%%%%%%%%%%%%%%%%%%%%%%%
We define three traffic scenarios. The first scenario is a hypothetical scenario, in which the road has a constant gradient $G(x) = -0.005$ for any position $x$~\citep{Goni-Ros2016}. We do not activate any JAD strategies in this scenario.
In the other scenarios, the road has a sag, and the road gradient distribution is given by Eq.~(\ref{eq:G_x}).
The second scenario is a baseline scenario, in which no JAD strategy is activated. We use the first and second scenarios to check the influence of the sag on the performance indicators defined in Sec.~\ref{sec:performance_indicators}.
In the third scenario, we activate a simple JAD strategy designed as follows.
%%%%%%%%%%%%%%%%%%%%%%%%%%%%%%%%%%%%%%%%%%%%%%%%%%%%%%%%%%%%%%%%%%%%%%%%%%%%%%%%%%%%%%%%%%%%%%%%%%%%%%%%%%%%%
\subsection{\label{subsec:JAD_strategy}JAD strategy}
%%%%%%%%%%%%%%%%%%%%%%%%%%%%%%%%%%%%%%%%%%%%%%%%%%%%%%%%%%%%%%%%%%%%%%%%%%%%%%%%%%%%%%%%%%%%%%%%%%%%%%%%%%%%%
Our JAD strategy is composed of the three steps: (i) estimation of the traffic jam fixed at the sag, (ii) activation or update of the JAD, and (iii) update of vehicle positions and velocities.
Every time step, we execute the three steps as given below.
%%%%%%%%%%%%%%%%%%%%%%%%%%%%%%%%%%%%%%%%%%%%%%%%%%%%%%%%%%%%%%%%%%%%%%%%%%%%%%%%%%%%%%%%%%%%%%%%%%%%%%%%%%%%%%%%
\subsubsection{\label{subsubsec:estimate_traffic_jam} Step (i) estimation of the traffic jam fixed at the sag}
%%%%%%%%%%%%%%%%%%%%%%%%%%%%%%%%%%%%%%%%%%%%%%%%%%%%%%%%%%%%%%%%%%%%%%%%%%%%%%%%%%%%%%%%%%%%%%%%%%%%%%%%%%%%%%%%
To estimate the traffic jam fixed at the sag, we define $t_{\rm R}(i)$ (s), $x_{\rm R}(i)$ (m) and $h_{\rm R}(i)$ (s) as the time, vehicle $i$'s position, and the time headway when it escapes from the traffic jam fixed at the sag, respectively.
Moreover, we define $i_{\rm R}(t')$ as the vehicle index of the last vehicle that has escaped from the traffic jam fixed at the sag by the current time $t'$.
We use $i_{\rm R}(t') = -1$ to represent the situation where no vehicle has escaped from the jam yet.
Since there is no traffic jam at $t = 0\,{\rm s}$, we set $i_{\rm R}(0) = -1$.
If the traffic jam has arisen and the first vehicle has escaped from it, $i_{\rm R}(t')$ is greater than or equal to zero.
In the following part of this section, we use $t'' (= t' - \Delta t)$ as the time one step before the current time $t'$, and $t''' (\le t' - 2 \Delta t)$ as the time two steps or more before $t'$.

We calculate $t_{\rm R}(i)$, $x_{\rm R}(i)$, $h_{\rm R}(i)$, and $i_{\rm R}(t')$ in two cases. The first case is the case such that no vehicle had escaped from the jam by $t''$ (that is, $i_{\rm R}(t'') = -1$). The second case is the case such that at least one vehicle had already escaped from the jam by $t''$ (that is, $i_{\rm R}(t'') \ge 0$).

In the first case ($i_{\rm R}(t'') = -1$), we search for the first vehicle that got caught in the jam between $t'''$ and $t''' + \Delta t$, and escaped from the jam between $t''$ and $t'$.
We identify that vehicle $i$ got caught in the jam between $t'''$ and $t''' + \Delta t$ if
\begin{align}
v_i(t''' + \Delta t) < v_{\rm cau} \le v_i(t'''),
\label{eq:get_caught_in_the_jam}
\end{align}
where $v_{\rm cau}$ (${\rm m}/{\rm s}$) is the threshold velocity.
We identify that vehicle $i$, which had been caught in the jam, escaped from the jam between $t''$ and $t'$ if
\begin{align}
v_i(t'') < v_{\rm esc} \le v_i(t') \; \land \; x_i(t') > x_{\rm esc},
\label{eq:escaping_from_the_jam_first_case}
\end{align}
where $v_{\rm esc}$ (${\rm m}/{\rm s}$) and $x_{\rm esc}$ (${\rm m}$) are the threshold velocity and position, respectively.
Since the downstream front of the traffic jam is generally fixed near the sag, we impose $x_i(t') > x_{\rm esc}$ for this identification.
If we identify that vehicle $i$ is the first vehicle satisfying both conditions~(\ref{eq:get_caught_in_the_jam}) and (\ref{eq:escaping_from_the_jam_first_case}), we obtain
\begin{align}
x_{\rm R}(i)  &= x_i(t'') + \dfrac{ v_{\rm esc}^2 - v_i(t'')^2 }{ 2 a_i(t'')}, \label{eq:x_R_i_first_case} \\
t_{\rm R}(i)  &= t'' + \dfrac{ v_{\rm esc} - v_i(t'') }{ a_i(t'') }, \label{eq:t_R_i_first_case} \\
h_{\rm R}(i)  &= h_{\rm R,\,pre}, \label{eq:h_R_i_first_case} \\
i_{\rm R}(t') &= i, \label{eq:i_R_tdash_first_case}
\end{align}
where $h_{\rm R,\,pre}$ (s) is the preset value of time headway in escaping from the traffic jam.

In the second case ($i_{\rm R}(t'') \ge 0$), we assume that the downstream front of the jam does not move; that is, $x_{\rm R}(i)$ is constant since the first vehicle's escape, and is given by Eq.~(\ref{eq:x_R_i_first_case}).
If $x_i(t')$ firstly becomes equal to or greater than the latest escaping position $x_{\rm R}(i_{\rm R}(t''))$:
\begin{align}
x_i(t'') < x_{\rm R}(i_{\rm R}(t'')) \le x_i(t'),
\label{eq:escaping_from_the_jam_second_case}
\end{align}
we identify that vehicle $i$ escaped from the jam between $t''$ and $t'$, and obtain
\begin{align}
x_{\rm R}(i)  &= x_{\rm R}(i_{\rm R}(t'')), \label{eq:x_R_i_second_case} \\
t_{\rm R}(i)  &= t'' + \dfrac{2 \left\{ x_{\rm R}(i) -  x_i(t'') \right\} }{v_i(t'') + \sqrt{v_i(t'')^2 + 2 a_i(t'') \left\{ x_{\rm R}(i) - x_i(t'') \right\} }}, \label{eq:t_R_i_second_case} \\
h_{\rm R}(i)  &= t_{\rm R}(i) - t_{\rm R}(i - 1), \label{eq:h_R_i_second_case} \\
i_{\rm R}(t') &= i. \label{eq:i_R_tdash_second_case}
\end{align}

We set $v_{\rm cau} = 15\,{\rm m}/{\rm s}$, $v_{\rm esc} = 28\,{\rm m}/{\rm s}$, $x_{\rm esc}=1600\,{\rm m}$, and $h_{\rm R,\,pre}=2\,{\rm s}$ as listed in Table~\ref{table:param}.
%%%%%%%%%%%%%%%%%%%%%%%%%%%%%%%%%%%%%%%%%%%%%%%%%%%%%%%%%%%%%%%%%%%%%%%%%%%%%%%%%%%%%%%%%%%%%%%%%%%%%%%%
\subsubsection{\label{subsubsec:activation_or_update_JAD} Step (ii) activation or update of the JAD}
%%%%%%%%%%%%%%%%%%%%%%%%%%%%%%%%%%%%%%%%%%%%%%%%%%%%%%%%%%%%%%%%%%%%%%%%%%%%%%%%%%%%%%%%%%%%%%%%%%%%%%%%
We execute step (ii) only if $i_{\rm R}(t') \ge 0$. Otherwise, we skip step (ii) and proceed to step (iii). We execute step (ii) in two cases. The first (second) case is the case such that the JAD is not activated (is activated).

In the first case, we choose the absorbing vehicle, and activate the JAD as follows.
We define
\begin{align}
x_{\rm JAD} = x_{\rm R}(i_{\rm R}(t')) - m (d + s_0 + T v_0)
\label{eq:x_JAD}
\end{align}
as the reference position to start the JAD.
Equation~(\ref{eq:x_JAD}) denotes that $x_{\rm JAD}$ is upstream of $x_{\rm R}(i_{\rm R}(t'))$ by the length of $m (d + s_0 + T v_0)$, which is the minimum required space for $m$ vehicles to run at the maximum velocity $v_0$.
Parameter $m$ ($m \in \mathbb{N}$ and $1 \le m \le N$) determines the spatiotemporal scale of the JAD. As $m$ increases, the spatiotemporal region for the JAD becomes wider.
If vehicle $i_{\rm JAD} \in [i_{\rm R}(t') + 1, N]$ satisfies
\begin{align}
x_{i_{\rm JAD}}(t') \le x_{\rm JAD} < x_{i_{\rm JAD} - 1}(t'),
\label{eq:i_JAD}
\end{align}
vehicle $i_{\rm JAD}$ is assigned to the absorbing vehicle, and starts the slow-in phase from the current time $t'$.
If we do not find vehicle index $i_{\rm JAD}$ satisfying condition~(\ref{eq:i_JAD}), we do not activate the JAD, skip the following procedure of step (ii), and proceed to step (iii).

In the slow-in phase, the absorbing vehicle moves toward the time-space point of the goal $(t_{\rm G}(t'), x_{\rm G}(t'))$, where $t_{\rm G}(t')$ (s) and $x_{\rm G}(t')$ (m) are the time and position of the goal, respectively.
Since we assume that the downstream front of the jam is fixed, $x_{\rm G}(t')$ is given by
\begin{align}
x_{\rm G}(t') = x_{\rm R}(i_{\rm R}(t')).
\label{eq:x_G}
\end{align}
To estimate $t_{\rm G}(t')$, we approximate that the time interval from when vehicle $i_{\rm R}(t')$ escapes from the jam to when the absorbing vehicle (vehicle $i_{\rm JAD}$) reaches $x_{\rm G}(t')$ is $\left\{ i_{\rm JAD} - i_{\rm R}(t') \right\} h_{\rm R}(i_{\rm R}(t'))$.
Besides, we limit the range of $h_{\rm R}(i_{\rm R}(t'))$ to $[h_{\rm R,\,min}, h_{\rm R,\,max}]$, where $h_{\rm R,\,min}$ (s) and $h_{\rm R,\,max}$ (s) are the minimum and maximum preset values for $h_{\rm R}(i_{\rm R}(t'))$, respectively.
We estimate $t_{\rm G}(t')$ as the time when this time interval elapses from $t_{\rm R}(i_{\rm R}(t'))$:
\begin{align}
t_{\rm G}(t') &= t_{\rm R}(i_{\rm R}(t')) + \left\{ i_{\rm JAD} - i_{\rm R}(t') \right\} \min \left\{ h_{\rm R,\,max},\;\max \left\{ h_{\rm R,\,min},\;h_{\rm R}(i_{\rm R}(t')) \right\} \right\}.
\label{eq:t_G}
\end{align}
In this way, we set the goal $(t_{\rm G}(t'), x_{\rm G}(t'))$ near the future time-space point where the DFTJ disappears.

We define $a_{\rm JAD}(t')$ (${\rm m}/{\rm s^2}$) as the acceleration designed for the slow-in phase.
To set $a_{\rm JAD}(t')$, we define $v_{\rm JAD, const}(t')$ (${\rm m}/{\rm s}$) as the constant velocity required by the absorbing vehicle to transit from the current time-space point $(t', x_{i_{\rm JAD}}(t'))$ to the goal $(t_{\rm G}(t'), x_{\rm G}(t'))$.
Velocity $v_{\rm JAD, const}(t')$ is given by
\begin{subequations}
\begin{empheq}[left={v_{\rm JAD, const}(t') = \empheqlbrace \,}]{align}
&v_0, & \quad \mbox{if $x_{\rm G}(t') < x_{i_{\rm JAD}}(t') \; \lor \; t_{\rm G}(t') - t' < 10^{-6}$,} \label{eq:v_JAD_const_a} \\
&\dfrac{x_{\rm G}(t') - x_{i_{\rm JAD}}(t')}{t_{\rm G}(t') - t'}, & \quad \mbox{otherwise,} \label{eq:v_JAD_const_b}
\end{empheq}
\label{eq:v_JAD_const}
\end{subequations}
where we use Eq.~(\ref{eq:v_JAD_const_a}) for preventing $v_{\rm JAD, const}(t')$ from becoming an unrealistic value.
We set $a_{\rm JAD}(t')$ so that the absorbing vehicle's velocity at the next time $v_{i_{\rm JAD}}(t' + \Delta t)$ approaches $v_{\rm JAD, const}(t')$ as follows:
\begin{subequations}
\begin{empheq}[left={a_{\rm JAD}(t') = \empheqlbrace \,}]{align}
&\max \left\{ -\dfrac{v_{i_{\rm JAD}}(t') - v_{\rm JAD, const}(t')}{\Delta t},\;a_{\rm min,\,JAD},\;-\dfrac{v_{i_{\rm JAD}}(t')}{\Delta t} \right\} & \quad \mbox{if $v_{i_{\rm JAD}}(t') > v_{\rm JAD, const}(t')$,} \label{eq:a_JAD_a} \\
&\min \left\{ \dfrac{v_{\rm JAD, const}(t') - v_{i_{\rm JAD}}(t')}{\Delta t},\;a_{\rm max,\,JAD},\;\dfrac{v_0 - v_{i_{\rm JAD}}(t')}{\Delta t} \right\} & \quad \mbox{if $v_{i_{\rm JAD}}(t') \le v_{\rm JAD, const}(t')$.} \label{eq:a_JAD_b}
\end{empheq}
\label{eq:a_JAD}
\end{subequations}
Parameters $a_{\rm min,\,JAD}$ (${\rm m}/{\rm s^2}$) and $a_{\rm max,\,JAD}$ (${\rm m}/{\rm s^2}$) are the minimum and maximum accelerations for the slow-in phase, respectively.
We use $-v_{i_{\rm JAD}}(t')/\Delta t$ and $(v_0 - v_{i_{\rm JAD}}(t'))/\Delta t$ in Eq.~(\ref{eq:a_JAD}) to restrict the absorbing vehicle's velocity from going below zero and beyond $v_0$, respectively.

In the second case, the absorbing vehicle (vehicle $i_{\rm JAD}$) is allowed to keep performing the action of the slow-in. The values of $x_{\rm G}(t')$, $t_{\rm G}(t')$ and $a_{\rm JAD}(t')$ are updated according to Eqs.~(\ref{eq:x_G}), (\ref{eq:t_G}) and (\ref{eq:a_JAD}), respectively.

We set $h_{\rm R,\,min} = 1.3\,{\rm s}$, $h_{\rm R,\,max}=2.5\,{\rm s}$, $a_{\rm min,\,JAD} = -1\,{\rm m}/{\rm s^2}$, and $a_{\rm max,\,JAD} = 1\,{\rm m}/{\rm s^2}$ as listed in Table~\ref{table:param}.
%%%%%%%%%%%%%%%%%%%%%%%%%%%%%%%%%%%%%%%%%%%%%%%%%%%%%%%%%%%%%%%%%%%%%%%%%%%%%%%%%%%%%%%%%%%%%%%%%%%%%%%%%%%%%
\subsubsection{\label{subsubsec:update_pos_vel} Step (iii) update of vehicle positions and velocities}
%%%%%%%%%%%%%%%%%%%%%%%%%%%%%%%%%%%%%%%%%%%%%%%%%%%%%%%%%%%%%%%%%%%%%%%%%%%%%%%%%%%%%%%%%%%%%%%%%%%%%%%%%%%%%
If the JAD is activated, the absorbing vehicle's actual acceleration $a_{i_{\rm JAD}}(t')$ is given by
\begin{align}
a_{i_{\rm JAD}}(t') = \max \left\{ \min \left\{ a_{\rm JAD}(t'),\;a_{{\rm des},\;i_{\rm JAD}}(t') \right\},\;a_{\rm min},\;-\dfrac{v_{i_{\rm JAD}}(t')}{\Delta t} \right\}.
\label{eq:a_i_JAD_t_dash}
\end{align}
The absorbing vehicle performs the action of the slow-in only if $a_{\rm JAD}(t')$ is greater than or equal to $a_{{\rm des},\,i_{\rm JAD}}(t')$. Otherwise, it suspends the slow-in phase from $t'$ to $t' + \Delta t$, and follows the preceding vehicle according to the IDM+ to prevent collisions.
It should be noted that we obtain $a_{i_{\rm JAD}}(t')$ by calculating Eq.~(\ref{eq:a_i_JAD_t_dash}) at each time step. Therefore, even if the absorbing vehicle suspends the slow-in phase at $t'$, it restarts the slow-in phase at $t' + \Delta t$ if $a_{\rm JAD}(t' + \Delta t) \ge a_{{\rm des},\,i_{\rm JAD}}(t' + \Delta t)$.
We use $a_{\rm min}$ and $-v_{i_{\rm JAD}}(t') / \Delta t$ in Eq.~(\ref{eq:a_i_JAD_t_dash}) as in Eq.~(\ref{eq:a_i_t}).
The accelerations of the other $N - 1$ vehicles are given by Eq.~(\ref{eq:a_i_t}).
If the JAD is not activated, the accelerations of all $N$ vehicles are given by Eq.~(\ref{eq:a_i_t}).

Subsequently, we update the positions and velocities of all vehicles according to the ballistic method as described in Eqs.~(\ref{eq:x_update_ballistic}) and (\ref{eq:v_update_ballistic}), respectively.

Next, if the JAD is activated and the absorbing vehicle becomes downstream of the fixed goal position at $t' + \Delta t$:
\begin{align}
x_{i_{\rm JAD}}(t' + \Delta t) > x_{\rm G}(t'),
\label{cond:slow_in_complete}
\end{align}
we judge that the absorbing vehicle has passed the goal between $t'$ and $t' + \Delta t$. In this case, the absorbing vehicle completes the slow-in phase (that is, finishes obeying Eq.~(\ref{eq:a_i_JAD_t_dash})) and starts the fast-out phase by following its preceding vehicle according to Eq.~(\ref{eq:a_i_t}) from $t' + \Delta t$.
Even when the absorbing vehicle completes its slow-in phase, we keep $i_{\rm R}(t' + \Delta t)$ at the latest value (larger than or equal to zero) without resetting $i_{\rm R}(t' + \Delta t) = -1$. At $t' + \Delta t$, we try to find the next absorbing vehicle, and assign it to start the next slow-in phase to remove the next traffic jam.
%%%%%%%%%%%%%%%%%%%%%%%%%%%%%%%%%%%%%%%%%%%%%%%%%%%%%%%%%%%%%%%%%%%%%%%%%%%%%%%%%%%%%%%%%
\section{\label{sec:performance_indicators}Performance indicators}
%%%%%%%%%%%%%%%%%%%%%%%%%%%%%%%%%%%%%%%%%%%%%%%%%%%%%%%%%%%%%%%%%%%%%%%%%%%%%%%%%%%%%%%%%
In this section, we define the performance indicators with respect to the travel time and the fuel consumption.
Travel time $T_i$ (s) is the length of time from the initial time $t = 0\,{\rm s}$ to the time when vehicle $i$ reaches the measurement end position $x_{\rm end}$ (m).
Total travel time $T_{\rm tot}$ (s) is the sum of $T_i$ for all vehicles:
\begin{align}
T_{\rm tot} = \sum_{i=1}^{N} T_i.
\label{eq:T_tot}
\end{align}
Total travel time $T_{\rm tot}$ denotes the efficiency of the whole system in terms of the travel time.
We set $x_{\rm end} = 5000\,{\rm m}$~\citep{Goni-Ros2016} as listed in Table~\ref{table:param}. This value of $x_{\rm end}$ is large enough for vehicles to recover their velocities after escaping from the traffic jam fixed at the sag.

Fuel consumption $F_i$ (kg) is the total amount of fuel consumed by vehicle $i$ to move from its initial position $x_{i}(0)$ to $x_{\rm end}$.
Total fuel consumption $F_{\rm tot}$ (kg) is the sum of $F_i$ for all vehicles:
\begin{align}
F_{\rm tot} = \sum_{i=1}^{N} F_i.
\label{eq:F_tot}
\end{align}
Total fuel consumption $F_{\rm tot}$ denotes the efficiency of the whole system in terms of the fuel consumption.
Additionally, we define $F_{{\rm cum},\,i}(t)$ (kg) as the cumulative amount of fuel consumed by vehicle $i$ from the initial time to time $t$.

To estimate the fuel consumption of each vehicle, we use the EMIT~\citep{Cappiello2002}. The EMIT determines the total tractive power requirement at the wheels of a vehicle at position $x$ represented as $P$ (kW) as follows:
\begin{align}
P = \max \left\{ 0,\; A v_{\rm mps} + B v_{\rm mps}^2 + C v_{\rm mps}^3 + M_{\rm ton} a_{\rm mpss} v_{\rm mps} + M_{\rm ton} g v_{\rm mps} \sin \left( \arctan G(x) \right) \right\},
\label{eq:P_T}
\end{align}
where $A$ (${\rm kW}/({\rm m}/{\rm s})$), $B$ (${\rm kW}/({\rm m}/{\rm s})^2$), and $C$ (${\rm kW}/({\rm m}/{\rm s})^3$) are the coefficients of the rolling resistance, the velocity-correction to the rolling resistance, and the air drag resistance, respectively, $M_{\rm ton}$ ($10^3 {\rm kg}$) is the vehicle mass, and $g$ (${\rm m}/{\rm s}^2$) is the gravitational constant.
Parameters $v_{\rm mps}$ (${\rm m}/{\rm s}$) and $a_{\rm mpss}$ (${\rm m}/{\rm s}^2$) are vehicle velocity and acceleration, respectively.
According to the value of $P$, the EMIT determines the instantaneous fuel consumption rate for the vehicle represented as $F_{\rm R}$ (${\rm g}/{\rm s}$) as follows:
\begin{subequations}
\begin{empheq}[left={F_{\rm R} = \empheqlbrace \,}]{alignat=2}
& \alpha_{\rm FR} + \beta_{\rm FR} v_{\rm kph} + \gamma_{\rm FR} v_{\rm kph}^2 + \delta_{\rm FR} v_{\rm kph}^3 + \zeta_{\rm FR} a_{\rm mpss} v_{\rm mps} & \quad \mbox{if $P > 10^{-6}$,} \label{eq:fuel_consumption_rate_a} \\
& \alpha'_{\rm FR} & \quad \mbox{if $P \le 10^{-6}$.} \label{eq:fuel_consumption_rate_b}
\end{empheq}
\label{eq:fuel_consumption_rate}
\end{subequations}
Parameter $v_{\rm kph}$ is vehicle velocity in the unit of ${\rm km}/{\rm h}$ (that is, $v_{\rm kph} = 3.6 v_{\rm mps}$).
Parameters $\alpha_{\rm FR}$ (${\rm g}/{\rm s}$), $\beta_{\rm FR}$ (${\rm g}/{\rm s}/({\rm km}/{\rm h})$), $\gamma_{\rm FR}$ (${\rm g}/{\rm s}/({\rm km}/{\rm h})^2$), $\delta_{\rm FR}$ (${\rm g}/{\rm s}/({\rm km}/{\rm h})^3$), $\zeta_{\rm FR}$ (${\rm g}/{\rm s}/({\rm m}^2/{\rm s}^3)$), and $\alpha'_{\rm FR}$ (${\rm g}/{\rm s}$) are coefficients.
Although multiple velocity units (${\rm km}/{\rm h}$ for $v_{\rm kph}$ and ${\rm m}/{\rm s}$ for $v_{\rm mps}$) appear in Eq.~(\ref{eq:fuel_consumption_rate}), we adhere to the units reported in Ref.~\citep{Cappiello2002}.
We obtain the amount of fuel consumption of each vehicle for every time interval of $\Delta t$ with the Euler method.

We follow the original setting of the EMIT parameters~\citep{Cappiello2002} for vehicle category 9 (light-duty vehicle, Tier 1 emission standards, cumulative mileage longer than 50000 miles, and high power-to-weight ratio)~\citep{Barth2000}.
This setting drops the term $\gamma_{\rm FR} v_{\rm kph}^2$ in Eq.~(\ref{eq:fuel_consumption_rate}).
All the EMIT parameters except for $\gamma_{\rm FR}$ are listed in Table~\ref{table:param}.
We also set the vehicle length $d=4.5\,{\rm m}$ as a typical vehicle length for category 9 as listed in Table~\ref{table:param}.

It should be noted that the EMIT was originally developed and calibrated for flat roads~\citep{Cappiello2002}. Nevertheless, we simply assume that the road gradient affects only $P$ through the term $M_{\rm ton} g v_{\rm mps} \sin \left( \arctan G(x) \right)$ in Eq.~(\ref{eq:P_T}); hence, we use the EMIT parameters for flat roads in our numerical simulations.

It also should be noted that Ref.~\citep{Cappiello2002} used conditions $P > 0$ and $P = 0$ for dividing Eq.~(\ref{eq:fuel_consumption_rate}) into Eqs.~(\ref{eq:fuel_consumption_rate_a}) and (\ref{eq:fuel_consumption_rate_b}), respectively. Since judging $P$ to be exactly zero may be severe, we use conditions $P > 10^{-6}$ and $P \le 10^{-6}$ instead for our numerical simulations.

To distinguish the hypothetical scenario, the baseline scenario, and the scenario with our JAD strategy, we add the subscripts ``hypo", ``base", and ``JAD" to performance indicators, respectively.
Additionally, when we emphasize that a performance indicator is obtained for given values of $N$ and $m$, we add subscripts $N$ and $m$ to the indicator, respectively.
For instance, $F_{{\rm tot,\,JAD},\,N,\,m}$ denotes the total fuel consumption for the scenario with our JAD strategy for given values of $N$ and $m$.

Moreover, we define the performance indicators of the system-size dependence with respect to the total travel time and the total fuel consumption. We define $\Delta T_{\rm tot,\,apv}(N,\,m)$ as the difference between the Average total Travel time per vehicle (AT) for the baseline scenario, and AT for the scenario with our JAD strategy as a function of $N$ and $m$:
\begin{align}
\Delta T_{\rm tot,\,apv}(N,\,m) = \dfrac{T_{{\rm tot,\,JAD},\,N,\,m} - T_{{\rm tot,\,base},\,N}}{N}.
\label{eq:Delta_T_tot_apv}
\end{align}
A negative (positive) value of $\Delta T_{\rm tot,\,apv}(N,\,m)$ denotes that our JAD strategy decreases (increases) AT from the baseline scenario.
Similarly, $\Delta F_{\rm tot,\,apv}(N,\,m)$ is the difference between the Average total Fuel consumption per vehicle (AF) for the baseline scenario, and AF for the scenario with our JAD strategy as a function of $N$ and $m$:
\begin{align}
\Delta F_{\rm tot,\,apv}(N,\,m) = \dfrac{F_{{\rm tot,\,JAD},\,N,\,m} - F_{{\rm tot,\,base},\,N}}{N}.
\label{eq:Delta_F_tot_apv}
\end{align}
A negative (positive) value of $\Delta F_{\rm tot,\,apv}(N,\,m)$ denotes that our JAD strategy decreases (increases) AF from the baseline scenario.
Additionally, we define $\Delta T_{\rm tot,\,apv,\,min}(N)$ and $\Delta F_{\rm tot,\,apv,\,min}(N)$ as the minimum values of $\Delta T_{\rm tot,\,apv}(N,\,m)$ and $\Delta F_{\rm tot,\,apv}(N,\,m)$, respectively for $m \in S(N)$:
\begin{align}
\Delta T_{\rm tot,\,apv,\,min}(N) &= \min_{m \in S(N)} \Delta T_{\rm tot,\,apv}(N,\,m),
\label{eq:Delta_T_tot_apv_min} \\
\Delta F_{\rm tot,\,apv,\,min}(N) &= \min_{m \in S(N)} \Delta F_{\rm tot,\,apv}(N,\,m),
\label{eq:Delta_F_tot_apv_min}
\end{align}
where $S(N)$ is a set of $m$ as a function of $N$, and is given in Sec.~\ref{subsubsec:system_size_dependence} (see Eq.~(\ref{eq:set_S_N})).
Values of $-\Delta T_{\rm tot,\,apv,\,min}(N)$ and $-\Delta F_{\rm tot,\,apv,\,min}(N)$ denote the Maximum reduction in the Average total Travel time per vehicle (MAT) and the Maximum reduction in the Average total Fuel consumption per vehicle (MAF) by our JAD strategy, respectively.
We define $m_{T_{{\rm tot}},\,{\rm opt}}(N)$ and $m_{F_{{\rm tot}},\,{\rm opt}}(N)$ as the values of $m$ to realize $\Delta T_{\rm tot,\,apv,\,min}(N)$ and $\Delta F_{\rm tot,\,apv,\,min}(N)$, respectively:
\begin{align}
m_{T_{{\rm tot}},\,{\rm opt}}(N) &= \arg \min_{m \in S(N)} \Delta T_{\rm tot,\,apv}(N,\,m),
\label{eq:m_T_tot_opt} \\
m_{F_{{\rm tot}},\,{\rm opt}}(N) &= \arg \min_{m \in S(N)} \Delta F_{\rm tot,\,apv}(N,\,m).
\label{eq:m_F_tot_opt}
\end{align}
Besides, we define $n_{T_{{\rm tot}},\,{\rm opt}}(N)$ and $n_{F_{{\rm tot}},\,{\rm opt}}(N)$ as the number of the actually assigned absorbing vehicles at $m = m_{T_{{\rm tot}},\,{\rm opt}}(N)$ and $m = m_{F_{{\rm tot}},\,{\rm opt}}(N)$, respectively in the scenario with our JAD strategy.
%%%%%%%%%%%%%%%%%%%%%%%%%%%%%%
\begin{figure}[t]
\centering
\includegraphics[width=\hsize]{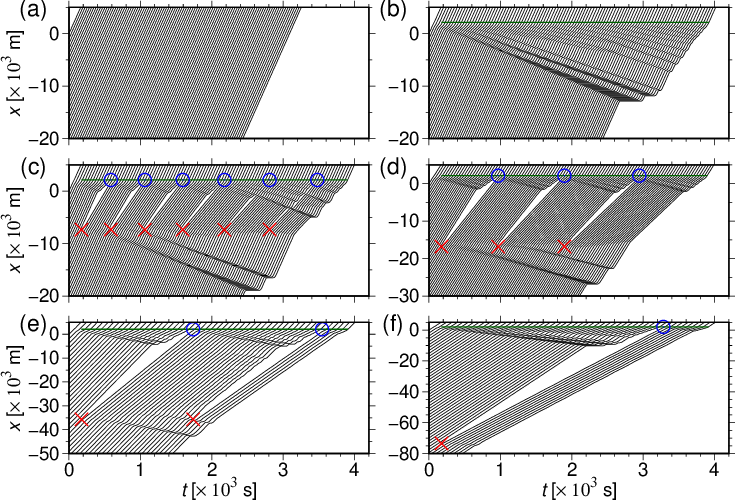}
\caption{
Time space diagrams for the system size $N=2000$. (a) The hypothetical scenario, in which the whole road is downhill. (b)--(f) Scenarios for the road with a sag. (b) The baseline scenario, in which we do not activate JAD. (c)--(f) The scenario with our JAD strategy. (c) $m=200$. (d) $m=400$. (e) $m=800$. (f) $m=1600$.
We depict the trajectories of (a)--(d) vehicles $1, 21, 41, \ldots, 1981$, and 2000 and (e)--(f) vehicles $1, 41, 81, \ldots, 1961$, and 2000.
Dark-green lines denote the downstream front of the traffic jam fixed at the sag detected by our JAD strategy.
Red crosses and blue circles denote the start and end points of the slow-in phase, respectively.
}
\label{fig:tx_N=2000}
\end{figure}
%%%%%%%%%%%%%%%%%%%%%%%%%%%%%%
\begin{figure}[htbp]
\centering
\includegraphics[width=\hsize]{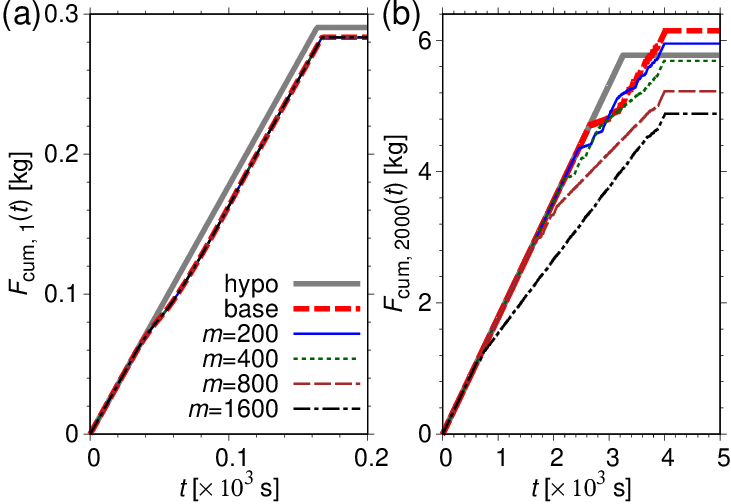}
\caption{
Cumulative fuel consumption of vehicle $i$ as a function of time $F_{{\rm cum}, i}(t)$ for $N=2000$. (a) $i = 1$. (b) $i = 2000$. Abbreviations hypo, base, and $m = 200$, 400, 800 and 1600 denote the hypothetical scenario, the baseline scenario, and the scenario with our JAD strategy for $m = 200$, 400, 800, and 1600, respectively.
}
\label{fig:F_cum_i_t_N=2000}
\end{figure}
%%%%%%%%%%%%%%%%%%%%%%%%%%%%%%
\begin{figure}[htbp]
\centering
\includegraphics[width=\hsize]{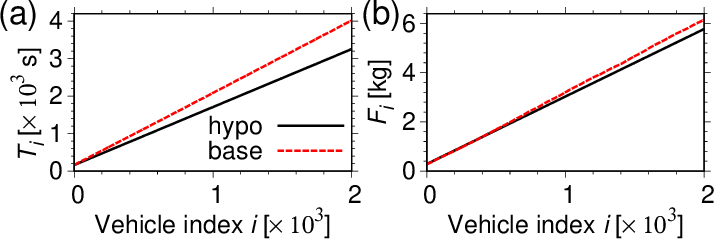}
\caption{
(a) Travel time $T_i$ and (b) fuel consumption $F_i$ as functions of the vehicle index $i$ ($i = 1, 2, \ldots, 2000$) for $N=2000$. Abbreviations hypo and base denote the hypothetical and the baseline scenarios, respectively.
}
\label{fig:T_i_F_i_N=2000}
\end{figure}
%%%%%%%%%%%%%%%%%%%%%%%%%%%%%%
%%%%%%%%%%%%%%%%%%%%%%%%%%%%%%%%%%%%%%%%%%%%%%%%%%%%%%%%%%%%%%%%%%%%%%%%%%%%%%%%%%%%%%%%%%%%%%%%%%%%%
\section{\label{sec:results}Results}
%%%%%%%%%%%%%%%%%%%%%%%%%%%%%%%%%%%%%%%%%%%%%%%%%%%%%%%%%%%%%%%%%%%%%%%%%%%%%%%%%%%%%%%%%%%%%%%%%%%%%
%%%%%%%%%%%%%%%%%%%%%%%%%%%%%%%%%%%%%%%%%%%%%%%%%%%%%%%%%%%%%%%%%%%%%%%%%%%%%%%%%%%%%%%%%%%%%%%%%%%%%%%%%%%%
\subsection{\label{subsec:sim_JAD_OFF}Scenarios without the activation of our JAD strategy}
%%%%%%%%%%%%%%%%%%%%%%%%%%%%%%%%%%%%%%%%%%%%%%%%%%%%%%%%%%%%%%%%%%%%%%%%%%%%%%%%%%%%%%%%%%%%%%%%%%%%%%%%%%%%
We conduct numerical simulations for the hypothetical and the baseline scenarios under $N = 2000$.
Figure~\ref{fig:tx_N=2000}(a) shows the time-space diagram for the hypothetical scenario, in which all the vehicles maintain their velocity at $v_0$ throughout the run, and no traffic jam occurs.
Figure~\ref{fig:tx_N=2000}(b) shows the time-space diagram for the baseline scenario, in which traffic jam occurs near the sag, and its downstream front is fixed. We depict the downstream front of the jam by a dark green line as a guide to the eyes as in Figs.~\ref{fig:tx_N=2000}(c)--(f) (see Sec.~\ref{subsubsec:sim_JAD_fixed_system_size}).
Multiple wide moving jams arise inside the jam, and propagate downstream. Vehicles that encounter multiple wide moving jams (such as vehicle $N$) decelerate and accelerate alternately before going through the sag.

% revised data:
% N_dependence_vol247_IDM-plus: N=2000, hypo
% N_dependence_vol248_IDM-plus: N=2000, base
%%%%%%%%%%%%%%%%%%%%%%
% #N	n_JAD	num_actual_JAD_vehicles	T_tot[s]	F_tot_SIDRA[mL]	F_tot_EMIT[g]
% 2000	2000	0	3416636.429057	6035893.555407	6063897.583838
% 2000	2000	0	4173671.774414	6657717.009917	6391532.500383
%%%%%%%%%%%%%%%%%%%%%%
% 4173671.774414 - 3416636.429057 = 757035.345357
% (4173671.774414 - 3416636.429057) / 3416636.429057 = 0.22157328152
%%%%%%%%%%%%%%%%%%%%%%
% 6391532.500383 - 6063897.583838 = 327634.916545
% (6391532.500383 - 6063897.583838) / 6063897.583838 = 0.05403041723
%%%%%%%%%%%%%%%%%%%%%%%%%%%%%%%%%%%%%%%%%%%%%%%%%%%%%%%%%%%%%%%%%%%%%%%%%%
% first submission data
% N_dependence_vol232_IDM-plus: N=2000, hypo
% N_dependence_vol233_IDM-plus: N=2000, base
%%%%%%%%%%%%%%%%%%%%%%
% #N	n_JAD	num_actual_JAD_vehicles	T_tot[s]	F_tot_SIDRA[mL]	F_tot_EMIT[g]
% 2000	2000	0	3416636.363636	6035893.439831	6063897.467725
% 2000	2000	0	4173671.774323	6657716.768091	6391532.248711
%%%%%%%%%%%%%%%%%%%%%%
% 4173671.774323 - 3416636.363636 = 757035.410687
% (4173671.774323 - 3416636.363636) / 3416636.363636 = 0.22157330488
%%%%%%%%%%%%%%%%%%%%%%
% 6391532.248711 - 6063897.467725 = 327634.780986
% (6391532.248711 - 6063897.467725) / 6063897.467725 = 0.05403039591
%%%%%%%%%%%%%%%%%%%%%%
We evaluate the influence of the sag on the total travel time and the total fuel consumption by comparing the two scenarios.
The obtained total travel times are $T_{\rm tot,\,hypo} = 3.417 \times 10^6\,{\rm s}$ and $T_{\rm tot,\,base} = 4.174 \times 10^6\,{\rm s}$; therefore, the sag increases the total travel time by $7.57 \times 10^5\,{\rm s}$ ($22.2\,{\rm \%}$).
The obtained total fuel consumptions are $F_{\rm tot,\,hypo} = 6.064 \times 10^3\,{\rm kg}$ and $F_{\rm tot,\,base} = 6.392 \times 10^3\,{\rm kg}$; therefore, the sag increases the total fuel consumption by $3.28 \times 10^2\,{\rm kg}$ ($5.40\,{\rm \%}$).

Figure~\ref{fig:F_cum_i_t_N=2000} shows the cumulative fuel consumption $F_{{\rm cum},\,i}(t)$ for (a) $i = 1$ and (b) $i = 2000$. Thick solid gray lines and thick dashed red lines denote the hypothetical scenario and the baseline scenario, respectively.
It should be noted that $F_{{\rm cum},\,i}(t)$ becomes stable (reaches $F_i$) after vehicle $i$ reaches the measurement end position $x_{\rm end}$.

In Fig.~\ref{fig:F_cum_i_t_N=2000}, the fuel consumption rate of each vehicle (the slope of $F_{{\rm cum},\,i}(t)$, that is, $F_{\rm R}$ for vehicle $i \in \left\{ 1, 2000 \right\}$) is constant until it reaches $x_{\rm end}$ in the hypothetical scenario. In contrast, the fuel consumption rate of each vehicle decreases in the baseline scenario because vehicles 1 and 2000 decelerate owing to the uphill and mainly the traffic jam, respectively.

In Fig.~\ref{fig:F_cum_i_t_N=2000}, $T_{\rm base,\,1}$ (the time when $F_{\rm cum,\,base,\,1}(t)$ becomes the final value $F_{\rm base,\,1}$) is slightly longer than $T_{\rm hypo,\,1}$ owing to the presence of uphill.
However, $T_{\rm base,\,2000}$ is significantly longer than $T_{\rm hypo,\,2000}$ because of the occurrence of the traffic jam (see also Figs.~\ref{fig:tx_N=2000}(a) and (b)).

In Fig.~\ref{fig:F_cum_i_t_N=2000}, $F_{\rm base,\,1}$ (the final value of $F_{\rm cum,\,base,\,1}(t)$) is smaller than $F_{\rm hypo,\,1}$ because the uphill decreases the fuel consumption rate of vehicle 1, and does not significantly prolong its travel time.
In contrast, $F_{\rm base,\,2000}$ is larger than $F_{\rm hypo,\,2000}$ because the traffic jam decreases the fuel consumption rate of vehicle 2000, \textit{but} significantly prolongs its travel time.

Figure~\ref{fig:T_i_F_i_N=2000}(a) shows $T_{{\rm base},\,i}$ and $T_{{\rm hypo},\,i}$ as functions of the vehicle index $i \in \left\{ 1, 2, \ldots, 2000 \right\}$. Vehicle $i$'s extra travel time caused by the sag, which is given by $T_{{\rm base},\,i} - T_{{\rm hypo},\,i}$, is positive and tends to increase monotonically as $i$ increases.
This monotonically increasing delay is caused by the extension of the traffic jam. Vehicle $i$ stays in the jam for longer periods as $i$ increases because the upstream tail of the jam propagates upstream, and the downstream front of the jam is fixed near the sag.

Figure~\ref{fig:T_i_F_i_N=2000}(b) shows $F_{{\rm base},\,i}$ and $F_{{\rm hypo},\,i}$ as functions of the vehicle index $i \in \left\{ 1, 2, \ldots, 2000 \right\}$. Vehicle $i$'s extra fuel consumption caused by the sag, which is given by $F_{{\rm base},\,i} - F_{{\rm hypo},\,i}$, also tends to increase monotonically as $i$ increases.
Running at a significantly low velocity brings a low fuel consumption rate \textit{but} an excessively long travel time to vehicles, and results in the increase in their fuel consumption.
%%%%%%%%%%%%%%%%%%%%%%%%%%%%%%%%%%%%%%%%%%%%%%%%%%%%%%%%%%%%%%%%%%%%%%%%%%%%%%%%%%%%%%%%%%%%%%%%%%%%%
\subsection{\label{subsec:stability}Stability analysis}
%%%%%%%%%%%%%%%%%%%%%%%%%%%%%%%%%%%%%%%%%%%%%%%%%%%%%%%%%%%%%%%%%%%%%%%%%%%%%%%%%%%%%%%%%%%%%%%%%%%%%
%%%%%%%%%%%%%%%%%%%%%%%%%%%%%%
\begin{figure}[t]
\centering
\includegraphics[width=\hsize]{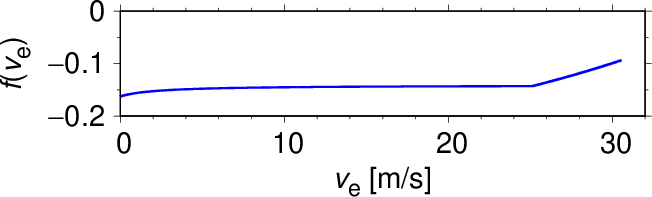}
\caption{
Function $f(v_{\rm e})$ for $0 \le v_{\rm e} \le v_0$ under the parameter settings listed in Table~\ref{table:param}.
}
\label{fig:f_ve}
\end{figure}
%%%%%%%%%%%%%%%%%%%%%%%%%%%%%%
To understand the reason for the occurrence of the traffic jam as shown in Fig.~\ref{fig:tx_N=2000}(b), we analyze the string stability for the traffic flow obeying the IDM+.
The string stability determines the growth or decay of a perturbation propagating through a vehicle platoon composed of an arbitrarily large number of vehicles~\citep{Treiber2013}. If the perturbation grows, the traffic flow is string unstable. Otherwise, the traffic flow is string stable.
Since the linear stability (stability against an infinitesimal perturbation) is generally more tractable than the nonlinear stability (stability against a finite perturbation), we consider the linear string stability.

We review the linear string stability condition for a general microscopic car-following model~\citep{Wilson2008b, Treiber2013}. This model expresses the acceleration of vehicle $i$ by $\tilde{a}_{\rm mic}(s_i(t), v_i(t), \Delta v_i(t))$, which is a function of the gap between vehicle $i$ and its preceding vehicle (vehicle $i-1$) $s_i(t)$, its velocity $v_i(t)$, and the relative velocity $\Delta v_i(t) = v_i(t) - v_{i-1}(t)$.
We consider a vehicle platoon composed of an infinite number of vehicles. Each vehicle in the platoon obeys the general model, and runs in equilibrium with a gap $s_{\rm e}$, equilibrium velocity $v_{\rm e}(s_{\rm e})$, and a relative velocity of zero.
This platoon is linearly string stable if the following condition is satisfied~\citep{Wilson2008b, Treiber2013}:
\begin{align}
\dfrac{{\rm d} v_{\rm e}(s_{\rm e})}{{\rm d} s} \le
- \dfrac{1}{2} \dfrac{\partial \tilde{a}_{\rm mic}(s_{\rm e}, v_{\rm e}, 0)}{\partial v_i}
- \dfrac{\partial \tilde{a}_{\rm mic}(s_{\rm e}, v_{\rm e}, 0)}{\partial \Delta v_i}.
\label{eq:stab_a_mic}
\end{align}
Otherwise, this platoon is linearly string unstable.
It should be noted that condition~(\ref{eq:stab_a_mic}) is only applicable for microscopic car-following models that satisfy the following two conditions:
\begin{align}
\dfrac{\partial \tilde{a}_{\rm mic}(s_{\rm e}, v_{\rm e}, 0)}{\partial v_i} &< 0
\label{eq:stab_a_mic_nec_cond_1}
\end{align}
and
\begin{align}
\dfrac{{\rm d} v_{\rm e}(s_{\rm e})}{{\rm d} s} &\ge 0.
\label{eq:stab_a_mic_nec_cond_2}
\end{align}

Since we consider the IDM+ as a specific car-following model, the equilibrium velocity $v_{\rm e}(s_{\rm e})$ is given by
\begin{subequations}
\begin{empheq}[left={v_{\rm e}(s_{\rm e}) = \empheqlbrace \,}]{alignat=2}
& \dfrac{s_{\rm e} - s_0}{T} & \quad \mbox{if $s_0 \le s_{\rm e} \le s_0 + T v_0$}, \label{eq:v_e_a} \\
& v_0                        & \quad \mbox{if $s_{\rm e} \ge s_0 + T v_0$}, \label{eq:v_e_b}
\end{empheq}
\label{eq:v_e}
\end{subequations}
and the three variables in condition~(\ref{eq:stab_a_mic}) are given by
\begin{subequations}
\begin{empheq}[left={\dfrac{{\rm d} v_{\rm e}(s_{\rm e})}{{\rm d} s} = \empheqlbrace \,}]{alignat=2}
& \dfrac{1}{T} & \quad \mbox{if $s_0 < s_{\rm e} < s_0 + T v_0$}, \label{eq:v_e_s_e_prime_a} \\
& 0            & \quad \mbox{if $s_{\rm e} > s_0 + T v_0$}, \label{eq:v_e_s_e_prime_b}
\end{empheq}
\label{eq:v_e_s_e_prime}
\end{subequations}
\begin{align}
\dfrac{\partial \tilde{a}_{\rm mic}(s_{\rm e}, v_{\rm e}, 0)}{\partial v_i} = -a \max \left\{ \dfrac{\delta v_{\rm e}^{\delta - 1}}{v_0^{\delta}},\,\dfrac{2 T (s_0 + T v_{\rm e})}{s_{\rm e}^2} \right\},
\label{eq:partial_a_mic_partial_v}
\end{align}
and
\begin{align}
\dfrac{\partial \tilde{a}_{\rm mic}(s_{\rm e}, v_{\rm e}, 0)}{\partial \Delta v_i} = - \dfrac{v_{\rm e} (s_0 + T v_{\rm e})}{s_{\rm e}^2} \sqrt{\dfrac{a}{b}}.
\label{eq:partial_a_mic_partial_delta_v}
\end{align}
Hence, the two necessary conditions~(\ref{eq:stab_a_mic_nec_cond_1}) and (\ref{eq:stab_a_mic_nec_cond_2}) for IDM+ are equivalent to
\begin{align}
s_0 < s_{\rm e} < s_0 + T v_0 \, \lor \, s_{\rm e} > s_0 + T v_0,
\label{eq:stab_nec_conds_IDM_plus}
\end{align}
and the linear string stability condition for IDM+ is given by
\begin{subequations}
\begin{empheq}[left={\empheqlbrace \,}]{alignat=2}
f(v_{\rm e}) \equiv \,
& \dfrac{a}{2} \max \left\{ \dfrac{\delta v_{\rm e}^{\delta - 1}}{v_0^{\delta}},\,\dfrac{2 T}{s_0 + T v_{\rm e}} \right\} + \dfrac{v_{\rm e}}{s_0 + T v_{\rm e}} \sqrt{\dfrac{a}{b}} - \dfrac{1}{T} \ge 0 & \quad \mbox{if $s_0 < s_{\rm e} < s_0 + T v_0$ ($\Leftrightarrow 0 < v_{\rm e} < v_0$)},
\label{eq:stab_IDM_plus_a} \\
& \dfrac{a}{2} \max \left\{ \dfrac{\delta}{v_0},\,\dfrac{2 T (s_0 + T v_0)}{s_{\rm e}^2} \right\} + \dfrac{v_0 (s_0 + T v_0)}{s_{\rm e}^2} \sqrt{\dfrac{a}{b}} \ge 0 & \quad \mbox{if $s_{\rm e} > s_0 + T v_0$},
\label{eq:stab_IDM_plus_b}
\end{empheq}
\label{eq:stab_IDM_plus}
\end{subequations}
where $f(v_{\rm e})$ is the left-hand side of condition~(\ref{eq:stab_IDM_plus_a}) as a function of $v_{\rm e}$.
Under the parameter settings listed in Table~\ref{table:param}, $f(v_{\rm e})$ is always negative for $0 < v_{\rm e} < v_0$ as shown in Fig.~\ref{fig:f_ve}, and the left-hand side of condition~(\ref{eq:stab_IDM_plus_b}) is always greater than zero for $s_{\rm e} > s_0 + T v_0$.
Accordingly, the traffic flow obeying the IDM+ under our parameter settings is linearly string unstable for $s_0 < s_{\rm e} < s_0 + T v_0$ ($\Leftrightarrow 0 < v_{\rm e} < v_0$), and linearly string stable for $s_{\rm e} > s_0 + T v_0$.

The traffic jam as shown in Fig.~\ref{fig:tx_N=2000}(b) occurs because the traffic flow is linearly string unstable for $0 < v_{\rm e} < v_0$, and the sag causes vehicles to decelerate from velocity $v_0$.
%%%%%%%%%%%%%%%%%%%%%%%%%%%%%%
\begin{figure}[htbp]
\centering
\includegraphics[width=\hsize]{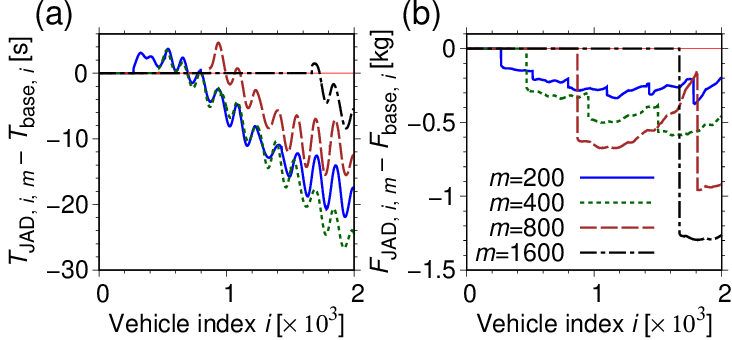}
\caption{
(a) Difference in vehicle $i$'s travel time between the baseline scenario and the scenario with our JAD strategy $T_{{\rm JAD},\,i,\,m} - T_{{\rm base},\,i}$ as a function of the vehicle index $i$.
(b) Difference in vehicle $i$'s fuel consumption between the two scenarios $F_{{\rm JAD},\,i,\,m} - F_{{\rm base},\,i}$ as a function of $i$.
We set $N=2000$ and $m \in \left\{ 200, 400, 800, 1600 \right\}$.
Red thin lines denote zero as guides to the eyes.
}
\label{fig:T_i_F_i_JAD_minus_base_N=2000}
\end{figure}
%%%%%%%%%%%%%%%%%%%%%%%%%%%%%%
%%%%%%%%%%%%%%%%%%%%%%%%%%%%%%
\begin{figure}[htbp]
\centering
\includegraphics[width=\hsize]{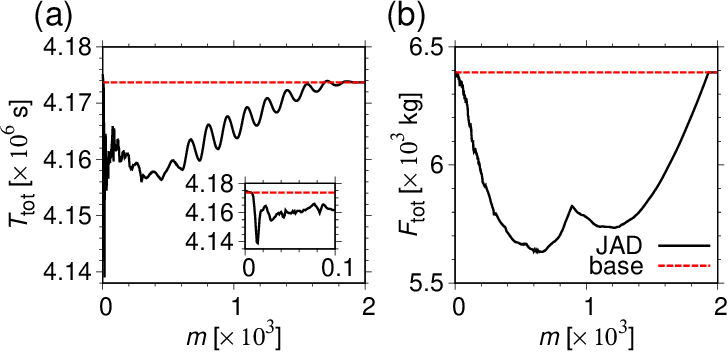}
\caption{
(a) Total travel time $T_{{\rm tot,\,JAD},\,m}$ as a function of $m$, and $T_{\rm tot,\,base}$.
(b) Total fuel consumption $F_{{\rm tot,\,JAD},\,m}$ as a function of $m$, and $F_{\rm tot,\,base}$.
We set $N = 2000$ and $m = 1, 2, \ldots, 2000$.
Abbreviations base and JAD denote the baseline scenario and the scenario with our JAD strategy, respectively.
}
\label{fig:T_tot_F_tot_n_JAD_N=2000}
\end{figure}
%%%%%%%%%%%%%%%%%%%%%%%%%%%%%%
%%%%%%%%%%%%%%%%%%%%%%%%%%%%%%
\begin{figure}[htbp]
\centering
\includegraphics[width=\hsize]{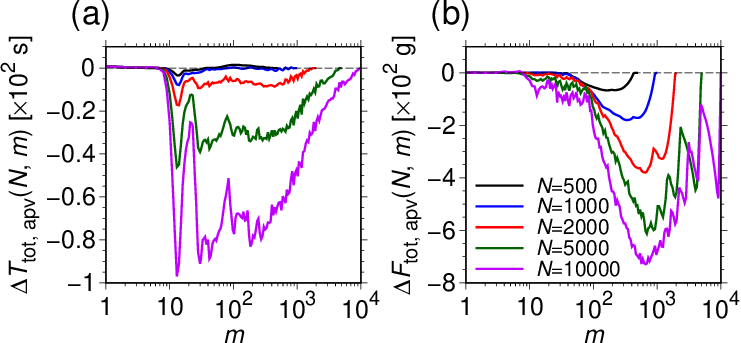}
\caption{
(a) The difference in average total travel times per vehicle between the baseline scenario and the scenario with our JAD strategy $\Delta T_{\rm tot,\,apv}(N,\,m)$.
(b) The difference in average total fuel consumptions per vehicle between the two scenarios $\Delta F_{\rm tot,\,apv}(N,\,m)$.
We set $N \in \left\{ 500, 1000, 2000, 5000, 10000 \right\}$ and $m \in S(N)$.
Thin dashed black lines denote zero as guides to the eyes.
}
\label{fig:Delta_T_tot_F_tot_apv_N_dependence}
\end{figure}
%%%%%%%%%%%%%%%%%%%%%%%%%%%%%%
%%%%%%%%%%%%%%%%%%%%%%%%%%%%%%%%%%%%%%%%%%%%%%%%%%%%%%%%%%%%%%%%%%%%%%%%%%%%%%%%%%%%%%%%%%%%%%%%%%%%%%%%%%%%%
\subsection{\label{subsec:sim_JAD}Scenario with our JAD strategy}
%%%%%%%%%%%%%%%%%%%%%%%%%%%%%%%%%%%%%%%%%%%%%%%%%%%%%%%%%%%%%%%%%%%%%%%%%%%%%%%%%%%%%%%%%%%%%%%%%%%%%%%%%%%%%
%%%%%%%%%%%%%%%%%%%%%%%%%%%%%%%%%%%%%%%%%%%%%%%%%%%%%%%%%%%%%%%%%%%%%%%%%%%%%%%%%%%%%%%%%%%%%%%%%%%%%%%%
\subsubsection{\label{subsubsec:sim_JAD_fixed_system_size}Results under a fixed system size}
%%%%%%%%%%%%%%%%%%%%%%%%%%%%%%%%%%%%%%%%%%%%%%%%%%%%%%%%%%%%%%%%%%%%%%%%%%%%%%%%%%%%%%%%%%%%%%%%%%%%%%%%
We check the basic characteristics of the scenario with our JAD strategy through numerical simulations under a fixed system size $N = 2000$.
Figures~\ref{fig:tx_N=2000}(c)--(f) show the obtained time-space diagrams for $m = 200$, $400$, $800$, and $1600$, respectively.
In each figure, the start and end points of the slow-in phase are depicted by red crosses and blue circles, respectively. Additionally, the downstream front of the traffic jam fixed at the sag is detected by our JAD strategy, and depicted by a dark green line. These figures indicate that the downstream front of the jam is detected successfully.

As $m$ increases, the target traffic jam propagates more upstream and the number of the absorbing vehicles decreases: six ($m = 200$), three ($m = 400$), two ($m = 800$), and one ($m = 1600$). In the case of $m = 200$, the fifth and sixth absorbing vehicles get involved in the target traffic jam during the slow-in phase because our JAD strategy does not consider the spatiotemporal trajectories of the vehicles just ahead of the absorbing vehicles unlike Ref.~\citep{Ghiasi2019}. In contrast to the case of $m = 200$, each absorbing vehicle does not get caught in the target traffic jam for $m \in \left\{400,\,800,\,1600 \right\}$.

Additionally, our JAD strategy causes the secondary traffic jams, which are wide moving jams, and occur upstream of the absorbing vehicles. The occurrence of the secondary traffic jams is inevitable because absorbing vehicles in the slow-in phase make the traffic flow upstream of them slower than $v_0$, and such traffic flow is linearly string unstable as discussed in Sec.~\ref{subsec:stability}.

Figures~\ref{fig:F_cum_i_t_N=2000}(a) and (b) show the cumulative fuel consumptions of the leading and the last vehicles under the scenario with our JAD strategy $F_{{\rm cum,\,JAD},\,1,\,m}(t)$ and $F_{{\rm cum,\,JAD},\,2000,\,m}(t)$, respectively for $m \in \left\{ 200, 400, 800, 1600 \right\}$.
For each $m$, $F_{{\rm cum,\,JAD},\,1,\,m}(t)$ coincides with $F_{{\rm cum,\,base},\,1}(t)$. In contrast, the final value of $F_{{\rm cum,\,JAD},\,2000,\,m}(t)$ is smaller than the final value of $F_{{\rm cum,\,base},\,2000}(t)$ for each $m$, and decreases with respect to $m$.

Figure~\ref{fig:T_i_F_i_JAD_minus_base_N=2000}(a) shows the difference in the travel time of vehicle $i$ between the baseline scenario and the scenario with our JAD strategy $T_{{\rm JAD},\,i,\,m} - T_{{\rm base},\,i}$ as a function of the vehicle index $i$ for $m \in \left\{ 200, 400, 800, 1600 \right\}$.
For each $m$, $T_{{\rm JAD},\,i,\,m} - T_{{\rm base},\,i}$ becomes roughly negative, and decreases roughly monotonically with respect to $i$ with fluctuation for the vehicles upstream of the first absorbing vehicle.

Figure~\ref{fig:T_i_F_i_JAD_minus_base_N=2000}(b) shows the difference in the fuel consumption of vehicle $i$ between the two scenarios $F_{{\rm JAD},\,i,\,m} - F_{{\rm base},\,i}$ as a function of the vehicle index $i$ for $m \in \left\{ 200, 400, 800, 1600 \right\}$.
For each $m$, $F_{{\rm JAD},\,i,\,m} - F_{{\rm base},\,i}$ decreases sharply after the appearance of each absorbing vehicle. The larger $m$, the greater the decrease.
After the sharp decrease, some of the fuel differences (in particular, $F_{{\rm JAD},\,i,\,800} - F_{{\rm base},\,i}$) increase gradually until the appearance of the next absorbing vehicle. The gradual increase occurs because vehicles between two successive absorbing vehicles get involved in the growing traffic jam fixed at the sag, and some of them also get involved in the growing secondary traffic jams.

% revised data
% N_dependence_vol246_IDM-plus
% #N	n_JAD	num_actual_JAD_vehicles	T_tot[s]	F_tot_SIDRA[mL]	F_tot_EMIT[g]
% T_tot_min: 2000	14	115	4139027.979809	6640851.449817	6374851.365672
% F_tot_min: 2000	657	2	4163008.348824	5820205.069636	5631930.715377
% base     : 2000	2000	0	4173671.774414	6657717.009917	6391532.500383
%%%%%%%%%%%%
% 4173671.774414 - 4139027.979809 = 34643.794605
% (4173671.774414 - 4139027.979809) / 4173671.774414 = 0.00830055559
%%%%%%%%%%%%
% 6391532.500383 - 5631930.715377 = 759601.785006
% (6391532.500383 - 5631930.715377) / 6391532.500383 = 0.11884501642
%%%%%%%%%%%%%%%%%%%%%%%%%%%%%%%%%%%%%%%%%%%%%%%%%%%%%%%%%%%%%%%%%%%%%%%%%%%%%%
% first submission data
% N_dependence_vol237_IDM-plus
% #N	n_JAD	num_actual_JAD_vehicles	T_tot[s]	F_tot_SIDRA[mL]	F_tot_EMIT[g]
% T_tot_min: 2000	14	115	4139027.979754	6640851.211646	6374851.215715
% F_tot_min: 2000	657	2	4163008.349159	5820204.803487	5631930.642436
% base     : 2000	2000	0	4173671.774323	6657716.768091	6391532.248711
%%%%%%%%%%
% 4173671.774323 - 4139027.979754 = 34643.794569
% (4173671.774323 - 4139027.979754) / 4173671.774323 = 0.00830055558
%%%%%%%%%%
% 6391532.248711 - 5631930.642436 = 759601.606275
% (6391532.248711 - 5631930.642436) / 6391532.248711 = 0.11884499314
%%%%%%%%%%
Figure~\ref{fig:T_tot_F_tot_n_JAD_N=2000}(a) shows the total travel time $T_{{\rm tot,\,JAD},\,m}$ as a function of $m$ ($m = 1, 2, \ldots, 2000$) with a solid black line, and $T_{\rm tot,\,base} = 4.174 \times 10^6\,{\rm s}$ with a dashed red line.
The total travel time $T_{{\rm tot,\,JAD},\,m}$ is smaller than $T_{\rm tot,\,base}$ for most $m$, and has the minimum value of $4.139 \times 10^6\,{\rm s}$ at $m = 14$. The maximum improvement in $T_{\rm tot}$ from the baseline scenario is $3.5 \times 10^4\,{\rm s}$ ($0.83\,{\rm \%}$).

Figure~\ref{fig:T_tot_F_tot_n_JAD_N=2000}(b) shows the total fuel consumption $F_{{\rm tot,\,JAD},\,m}$ as a function of $m$ ($m = 1, 2, \ldots, 2000$) with a solid black line, and $F_{\rm tot,\,base} = 6.392 \times 10^3\,{\rm kg}$ with a dashed red line.
The total fuel consumption $F_{{\rm tot,\,JAD},\,m}$ is smaller than $F_{\rm tot,\,base}$ for most $m$, roughly has two local minimum values, and has the minimum value of $5.632 \times 10^3\,{\rm kg}$ at $m = 657$. The maximum improvement in $F_{\rm tot}$ from the baseline scenario is $7.60 \times 10^2\,{\rm kg}$ ($11.9\,{\rm \%}$).
%%%%%%%%%%%%%%%%%%%%%%%%%%%%%%
\begin{figure}[t]
\centering
\includegraphics[width=\hsize]{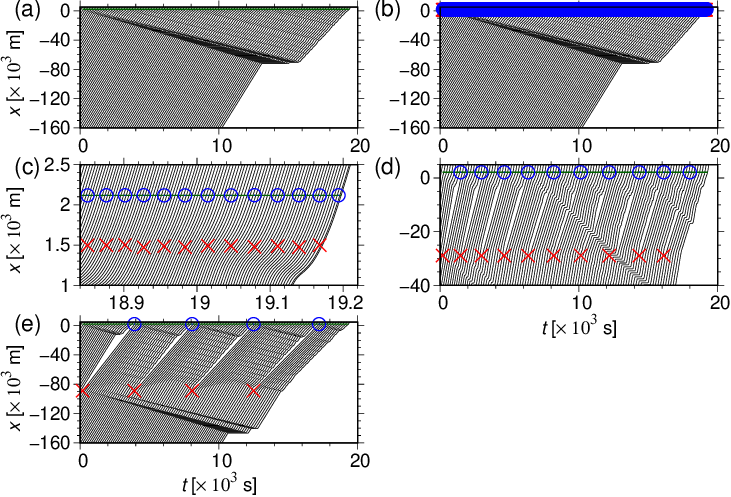}
\caption{
Time-space diagrams for $N = 10000$.
(a) The baseline scenario. (b)--(e) The scenario with our JAD strategy. (b) $m = 13$, which is the value of $m$ minimizing the total travel time. (c) $m = 13$ (enlarged view). (d) $m = 657$, which is the value of $m$ minimizing the total fuel consumption. (e) $m = 1924$.
We depict the trajectories of vehicles $1, 101, 201, \ldots, 9901$, and 10000 for (a), (b), (d), and (e), and those of vehicles $10000, 9999, 9997, 9995, 9993, \ldots$ for (c).
Dark-green lines denote the downstream front of the traffic jam fixed at the sag detected by our JAD strategy.
Red crosses and blue circles denote the start and end points of the slow-in phase, respectively.
}
\label{fig:tx_N=10000}
\end{figure}
%%%%%%%%%%%%%%%%%%%%%%%%%%%%%%
%%%%%%%%%%%%%%%%%%%%%%%%%%%%%%
\begin{figure}[t]
\centering
\includegraphics[width=\hsize]{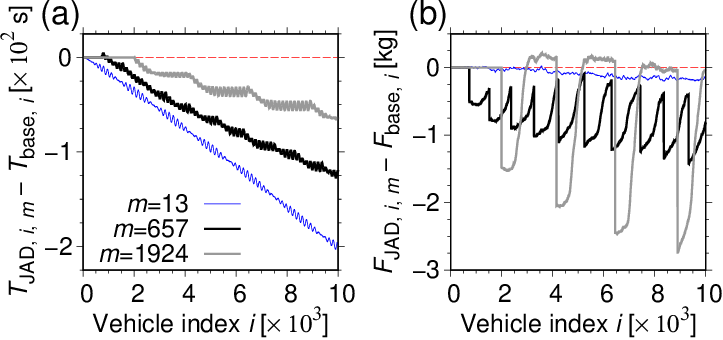}
\caption{
(a) $T_{{\rm JAD},\,i,\,m} - T_{{\rm base},\,i}$ and (b) $F_{{\rm JAD},\,i,\,m} - F_{{\rm base},\,i}$ as functions of the vehicle index $i$ for $N=10000$ and $m \in \left\{ 13, 657, 1924 \right\}$.
Thin red lines denote zero as guides to the eyes.
}
\label{fig:diff_T_tot_F_tot_from_base_vehicle_index_N=10000}
\end{figure}
%%%%%%%%%%%%%%%%%%%%%%%%%%%%%%
%%%%%%%%%%%%%%%%%%%%%%%%%%%%%%
\begin{figure}[htbp]
\centering
\includegraphics[width=\hsize]{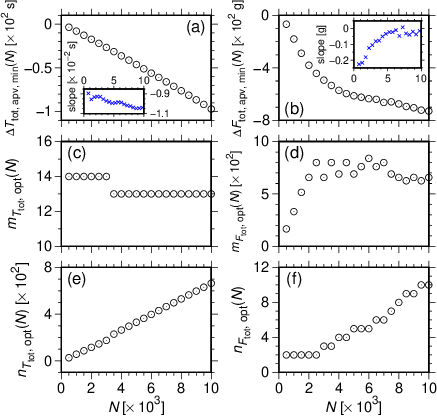}
\caption{
(a) $\Delta T_{\rm tot,\,\rm apv,\,min}(N)$, (b) $\Delta F_{\rm tot,\,\rm apv,\,min}(N)$, (c) $m_{T_{{\rm tot}},\,{\rm opt}}(N)$, (d) $m_{F_{{\rm tot}},\,{\rm opt}}(N)$, (e) $n_{T_{{\rm tot}},\,{\rm opt}}(N)$, and (f) $n_{F_{{\rm tot}},\,{\rm opt}}(N)$ for $N \in \left\{ 500, 1000, 1500, \ldots, 10000 \right\}$.
The insets in (a) and (b) show the slopes of $\Delta T_{\rm tot,\,\rm apv,\,min}(N)$ and $\Delta F_{\rm tot,\,\rm apv,\,min}(N)$, respectively.
}
\label{fig:min_values_N_dependence}
\end{figure}
%%%%%%%%%%%%%%%%%%%%%%%%%%%%%%
%%%%%%%%%%%%%%%%%%%%%%%%%%%%%%%%%%%%%%%%%%%%%%%%%%%%%%%%%%%%%%%%%%%%%%%%%%%%%%%%%%%%%%%%
\subsubsection{\label{subsubsec:system_size_dependence}System-size dependence}
%%%%%%%%%%%%%%%%%%%%%%%%%%%%%%%%%%%%%%%%%%%%%%%%%%%%%%%%%%%%%%%%%%%%%%%%%%%%%%%%%%%%%%%%
We investigate the influence of the system size $N$ on our JAD strategy through numerical simulations. We set $N \in \left\{ 500, 1000, 1500, \ldots, 10000 \right\}$. Since investigating all values of $m \in \left\{ 1, 2, \ldots, N \right\}$ requires large computational times for large $N$, we investigate all values of $m \in S(N)$ instead. Set $S(N)$ is given by
\begin{align}
S(N) = \left\{ \lfloor c^k \rfloor \,\middle|\, k = 0, 1, 2, \ldots, k_{\rm max}(N) \right\} \cup \left\{ N \right\},
\label{eq:set_S_N}
\end{align}
where $\lfloor \cdot \rfloor$ is the floor function, which returns the maximum integer less than or equal to the input real number (that is, for an input $r \in \mathbb{R}$, $\lfloor r \rfloor = \max \left\{ q \in \mathbb{Z} \mid q \le r \right\}$). We set $c = 1.05$ and $k_{\rm max}(N) = \lfloor \log_{10} N / \log_{10} c \rfloor$.
We believe that using $S(N)$ instead of $\left\{ 1, 2, \ldots, N \right\}$ is sufficient to reveal the influence of $m$ on output values.

Figure~\ref{fig:Delta_T_tot_F_tot_apv_N_dependence}(a) shows $\Delta T_{\rm tot,\,\rm apv}(N,\,m)$ for $N \in \left\{ 500, 1000, 2000, 5000, 10000 \right\}$ and $m \in S(N)$.
For roughly $m > 10$, $\Delta T_{\rm tot,\,\rm apv}(N,\,m)$ tends to decrease with respect to $N$. The minimum value of $\Delta T_{\rm tot,\,\rm apv}(10000,\,m)$ is $-97\,{\rm s}$ at $m = 13$.
Additionally, $\Delta T_{\rm tot,\,\rm apv}(N,\,m)$ fluctuates with respect to $m$ especially in the range of roughly $10 < m < 30$.

Figure~\ref{fig:Delta_T_tot_F_tot_apv_N_dependence}(b) shows $\Delta F_{\rm tot,\,\rm apv}(N,\,m)$ for $N \in \left\{ 500, 1000, 2000, 5000, 10000 \right\}$ and $m \in S(N)$.
For roughly $m > 10$, $\Delta F_{\rm tot,\,\rm apv}(N,\,m)$ roughly decreases with respect to $N$. The minimum value of $\Delta F_{\rm tot,\,\rm apv}(10000,\,m)$ is $-730\,{\rm g}$ at $m = 657$.
Additionally, $\Delta F_{\rm tot,\,\rm apv}(N,\,m)$ fluctuates with respect to $m$ especially for large values of $N$ and $m$, presumably owing to the small number of absorbing vehicles.

Figures~\ref{fig:tx_N=10000}(a)--(e) show the time-space diagrams for $N=10000$ in (a) the baseline scenario, and (b)--(e) the scenario with our JAD strategy for (b) $m = 13$ (minimizing $\Delta T_{{\rm tot},\,{\rm apv}}(N,\,m)$), (c) $m = 13$ (enlarged view), (d) $m = 657$ (minimizing $\Delta F_{{\rm tot},\,{\rm apv}}(N,\,m)$), and (e) $m = 1924$.
We depict dark-green lines, red crosses and blue circles in these diagrams as in Fig.~\ref{fig:tx_N=2000}.
In contrast to the cases of $m = 657$ and 1924, for $m = 13$, absorbing vehicles do not remove the target traffic jam because the starting positions of the slow-in phase are excessively close to the goal.

Figure~\ref{fig:diff_T_tot_F_tot_from_base_vehicle_index_N=10000}(a) shows the relationship between the vehicle index $i$ and $T_{{\rm JAD},\,i,\,m} - T_{{\rm base},\,i}$ for $N = 10000$ and $m \in \left\{ 13, 657, 1924 \right\}$.
For $m = 13$, $T_{{\rm JAD},\,i,\,13} - T_{{\rm base},\,i}$ decreases roughly constantly with fluctuation, and is the lowest among the three values of $m$ for most vehicle index $i$.
For $m = 657$ and $1924$, $T_{{\rm JAD},\,i,\,m} - T_{{\rm base},\,i}$ decreases stepwise. Especially, for $m = 1924$, the number of the stepwise decrease in $T_{{\rm JAD},\,i,\,1924} - T_{{\rm base},\,i}$ is four, which is equal to the number of the absorbing vehicles.

Figure~\ref{fig:diff_T_tot_F_tot_from_base_vehicle_index_N=10000}(b) shows the relationship between the vehicle index $i$ and $F_{{\rm JAD},\,i,\,m} - F_{{\rm base},\,i}$ for $N = 10000$ and $m \in \left\{ 13, 657, 1924 \right\}$.
For $m = 13$, $F_{{\rm JAD},\,i,\,13} - F_{{\rm base},\,i}$ decreases roughly constantly with fluctuation.
For $m = 657$ and $1924$, $F_{{\rm JAD},\,i,\,m} - F_{{\rm base},\,i}$ decreases sharply by the assignment of each absorbing vehicle, and increases till the assignment of the next absorbing vehicle as in Fig.~\ref{fig:T_i_F_i_JAD_minus_base_N=2000}(b).
For $m = 1924$, the decrease in $F_{{\rm JAD},\,i,\,1924} - F_{{\rm base},\,i}$ by one absorbing vehicle is the largest among the three values of $m$, while the increase in this value till the next absorbing vehicle is also the largest, and this value even becomes positive.

Figures~\ref{fig:min_values_N_dependence}(a), (c), and (e) show $\Delta T_{\rm tot,\,apv,\,min}(N)$, $m_{T_{\rm tot},\,{\rm opt}}(N)$, and $n_{T_{\rm tot},\,{\rm opt}}(N)$, respectively.
The inset of Fig.~\ref{fig:min_values_N_dependence}(a) shows the slope of $\Delta T_{\rm tot,\,apv,\,min}(N)$, which is given by $\{ \Delta T_{\rm tot,\,apv,\,min}(N_1 + 500) - \Delta T_{\rm tot,\,apv,\,min}(N_1) \} / 500$ at $N = N_1 + 250$ for $N_1 \in \{ 500, 1000, \ldots, 9500 \}$.
The output value $\Delta T_{\rm tot,\,apv,\,min}(N)$ decreases roughly linearly, and its slope tends to decrease slightly with respect to $N$. The optimal JAD scaling parameter $m_{T_{\rm tot},\,{\rm opt}}(N)$ is nearly constant at 13 or 14 with respect to $N$; therefore, our JAD strategy minimizes $T_{\rm tot}$ without removing the traffic jam for a wide range of the system size from $N = 500$ to $10000$.
The optimal number of absorbing vehicles $n_{T_{\rm tot},\,{\rm opt}}(N)$ tends to increase linearly with respect to $N$.

Figures~\ref{fig:min_values_N_dependence}(b), (d), and (f) show $\Delta F_{\rm tot,\,apv,\,min}(N)$, $m_{F_{\rm tot},\,{\rm opt}}(N)$, and $n_{F_{\rm tot},\,{\rm opt}}(N)$, respectively.
The inset of Fig.~\ref{fig:min_values_N_dependence}(b) shows the slope of $\Delta F_{\rm tot,\,apv,\,min}(N)$.
The output value $\Delta F_{\rm tot,\,apv,\,min}(N)$ decreases with respect to $N$. However, the rate of decrease in $\Delta F_{\rm tot,\,apv,\,min}(N)$ behaves differently from $\Delta T_{\rm tot,\,apv,\,min}(N)$. The slope of $\Delta F_{\rm tot,\,apv,\,min}(N)$ tends to increase with respect to $N$ till roughly $N = 5000$,
%%% first submission start %%%
% and become stable with fluctuation for larger $N$.
%%% first submission end %%%
%%% revised start %%%
and become constant with fluctuation for larger $N$.
%%% revised end %%%
The optimal number of absorbing vehicles $n_{F_{\rm tot},\,{\rm opt}}(N)$ is 2 for $N \le 2500$, and tends to increase roughly linearly with respect to $N$ for larger $N$.
Consequently, $m_{F_{\rm tot},\,{\rm opt}}(N)$ increases monotonically with $N$ for $N \le 2500$, and becomes constant with fluctuation for larger $N$.

Figures~\ref{fig:min_values_N_dependence}(c) and (d) reveal that the optimal value of $m$ becomes constant to minimize either AT or AF as long as $N$ is sufficiently large, \textit{but} we cannot optimize both of them simultaneously.
Nevertheless, using $m_{F_{\rm tot},\,{\rm opt}}(N)$ realizes a considerably large reduction in AT for large system sizes as shown in Fig.~\ref{fig:Delta_T_tot_F_tot_apv_N_dependence}; for instance, $65\,{\rm \%}$ of $\Delta T_{\rm tot,\,apv,\,min}(N)$ is achieved for $N = 10000$.
%%%%%%%%%%%%%%%%%%%%%%%%%%%%%%%%%%%%%%%%%%%%%%%%%%%%%%%%%%%%%%%%%%%%%
\section{\label{sec:discussion}Discussion}
%%%%%%%%%%%%%%%%%%%%%%%%%%%%%%%%%%%%%%%%%%%%%%%%%%%%%%%%%%%%%%%%%%%%%
We have considered CVs running on a single-lane road with a sag, designed a simple JAD strategy to remove the traffic jam caused by the sag, and conducted numerical simulations for system sizes ranging from $N = 500$ to $10000$ under the presence of traffic instability. We have found the following system-size dependence of our JAD strategy.
MAT increases roughly constantly, and the rate of increase in MAT increases slightly with respect to $N$. MAF increases with respect to $N$,
%%% first submission start %%%
% while the rate of increase in MAF roughly saturates with respect to $N$.
%%% first submission end %%%
%%% revised start %%%
while the rate of increase in MAF decreases and becomes roughly constant with respect to $N$.
%%% revised end %%%
As $N$ increases from 500 to 10000, both the optimal values of the JAD scaling parameter $m$ to realize MAT and MAF become constant; the former is much smaller than the latter, which denotes that our strategy can not minimize AT and AF simultaneously. Nevertheless, our JAD strategy can reduce AT considerably by minimizing AF for large $N$.

We have revealed that our JAD strategy reduces AT and AF for considerably large system sizes up to 10000 even though the strategy makes the upstream flow string unstable. Our results will stir further development of the strategies which manipulate the spatiotemporal maneuvers of special vehicles for large systems under the traffic instability.

Refs.~\citep{Piacentini2018a,Goni-Ros2016} reduced the total travel time without removing the traffic jam fixed at a bottleneck by deploying a single or a few slow vehicles. Our results also show that removing the traffic jam is not always necessary for reducing the total travel time in going through the sag for the system size ranging from $N = 500$ to $10000$. We guess that reducing the total travel time will rather require short time-headways after escaping from the traffic jam.

To focus on the system size effect, we have simplified the following conditions: the penetration rate of CVs ($100\,{\rm \%}$), vehicle types (cars only and no trucks), number of lanes (one), and number of sags (one). Additionally, we have not considered the effective range, delay, or error of the V2V or V2I communication. We have updated the acceleration designed for the slow-in phase $a_{\rm JAD}(t')$ every time step (at time intervals of $0.1\,{\rm s}$), and not at larger time intervals.
Varying these conditions will reveal the effectiveness of our JAD strategy in more severe or realistic traffic situations, which warrants our future work.
Moreover, we have not considered the traffic stochasticity~\citep{Wagner2012,Laval2014a,Jiang2014,Huang2018c}. Interestingly, traffic stochasticity was incorporated into a dynamic VSL strategy to eliminate a wide moving jam~\citep{Wang2012eIJMPC,Wang2014fIJMPC}. We will develop JAD strategies to mitigate traffic jam caused by a sag under the existence of the traffic stochasticity in future work.

Since we have constructed our JAD strategy as simply as possible, our JAD strategy has not prevented absorbing vehicles from getting involved in the target traffic jam unlike Ref.~\citep{Ghiasi2019}, or eliminated the secondary traffic jam unlike Refs.~\citep{He2017,Zheng2020}.
Moreover, our JAD strategy has used the maneuvers common to all absorbing vehicles unlike the different maneuvers for each special vehicle~\citep{Goni-Ros2016}. We will investigate the system-size dependence of more sophisticated JAD strategies in our future work.

We have not searched for the optimal maneuvers of the absorbing vehicles in terms of the total travel time or the total fuel consumption unlike the quasi-optimized maneuvers in terms of the total travel time~\citep{Goni-Ros2016}. Although it will need large computational costs, investigating the system-size dependence of the optimal trajectories of the absorbing vehicles will contribute to further development of JAD strategies, and warrants our future work.
%%%%%%%%%%%%%%%%%%%%%%%%%%%%%%%%%%%%%%%%%%%%%%%%%%%%%%%%%%%%%%%%%%%%%
\section*{Acknowledgements}
%%%%%%%%%%%%%%%%%%%%%%%%%%%%%%%%%%%%%%%%%%%%%%%%%%%%%%%%%%%%%%%%%%%%%
This work was supported by JSPS KAKENHI Grant Number JP20K03749.
%%%%%%%%%%%%%%%%%%%%%%%%%%%%%%%%%%%%%%%%%%%%%%%%%%%%%%%%%%%%%%%%%%%%%%%%%%%%%%%
%% The Appendices part is started with the command \appendix;
%% appendix sections are then done as normal sections
%% \appendix

%% \section{}
%% \label{}
%%%%%%%%%%%%%%%%%%%%%%%%%%%%%%%%%%%%%%%%%%%%%%%%%%%%%%%%%%%%%%%%%%%%%%%%%%%%%%%

%\clearpage
%%%%%%%%%%%%%%%%%%%%%%%%%%%%%%%%%%%%%%%%%%%%%%%%%%%%%%%%%%%%%%%%%%%%%%%%%%%%%%%
% references
%%%%%%%%%%%%%%%%%%%%%%%%%%%%%%%%%%%%%%%%%%%%%%%%%%%%%%%%%%%%%%%%%%%%%%%%%%%%%%%

% case 1: using bibtex
%%%%%%%%%%%%%%%%%%%%%%%%%%%%%%%%%%%%%%%%%%%%%
% \bibliographystyle{elsarticle-harv}
% \bibliographystyle{elsarticle-num}
% \bibliography{JAD_bn_bibtex}
%%%%%%%%%%%%%%%%%%%%%%%%%%%%%%%%%%%%%%%%%%%%%
% case 2: using bbl
%%%%%%%%%%%%%%%%%%%%%%%%%%%%%%%%%%%%%%%%%%%%%

\end{document}